\documentclass[reprint,prd,floatfix,
superscriptaddress,nofootinbib,notitlepage]{revtex4-1}
\pdfoutput=1
\usepackage{graphicx,epsfig}
\usepackage{amsmath}
\usepackage{amssymb}
\usepackage{longtable}
\usepackage{mathtools}
\usepackage{multirow}
\usepackage{comment}
\usepackage{dcolumn}
\usepackage{physics}
\usepackage{bm}
\usepackage{amsfonts}
\usepackage{subfigure}
\usepackage{color}
\usepackage{xcolor}
\usepackage{relsize}

\makeatletter
\newcommand\emailx[1]{%
\move@AF%
\def\@affil{{\normalfont\,#1\strut}{}}%
}%

\begin{document}

\title{Normal modes of Proca fields in AdS$_d$ spacetime}

\author{David Lopes}
\affiliation{Centro de Astrof\'{\i}sica e Gravita\c c\~ao  - CENTRA,
Departamento de F\'{\i}sica, Instituto Superior T\'ecnico - IST,
Universidade de Lisboa - UL, Avenida Rovisco Pais 1, 1049-001
Lisboa, Portugal\vskip 0.1cm}
\emailx{david.d.lopes@tecnico.ulisboa.pt}

\author{Tiago V. Fernandes}
\affiliation{Centro de Astrof\'{\i}sica e Gravita\c c\~ao  - CENTRA,
Departamento de F\'{\i}sica, Instituto Superior T\'ecnico - IST,
Universidade de Lisboa - UL, Avenida Rovisco Pais 1, 1049-001
Lisboa, Portugal\vskip 0.1cm}
\emailx{tiago.vasques.fernandes@tecnico.ulisboa.pt}

\author{Jos\'e P. S. Lemos}
\affiliation{Centro de Astrof\'{\i}sica e Gravita\c c\~ao  - CENTRA,
Departamento de F\'{\i}sica, Instituto Superior T\'ecnico - IST,
Universidade de Lisboa - UL, Avenida Rovisco Pais 1, 1049-001
Lisboa, Portugal\vskip 0.1cm}
\emailx{joselemos@ist.utl.pt}

\begin{abstract} 

The normal modes of Proca field perturbations in $d$-dimensional
anti-de Sitter spacetime, AdS$_d$ for short, with
reflective Dirichlet boundary
conditions, are obtained
exactly. Within the Ishibashi-Kodama framework, we decompose the Proca
field in scalar-type and vector-type components, according to their
tensorial behavior on the $(d-2)$-sphere $\mathcal{S}^{d-2}$.  Two of
the degrees of freedom of the Proca field are described by scalar-type
components, which in general are coupled due to the mass of the field,
but in AdS$_d$ we show that they can be decoupled.  The other $d-3$
degrees of freedom of the field are described by a vector-type
component that generically decouples completely.  The normal modes
and their frequencies for
both the scalar-type and vector-type components of the Proca field are
then obtained analytically. Additionally, we analyze the normal modes
of the Maxwell field as the massless limit of the Proca field.
We
find that for scalar-type perturbations
in $d=4$ there is a discontinuity in the massless limit,
in $d=5$ the massless limit
is well-defined using Dirichlet-Neumann rather than 
Dirichlet boundary conditions,
and in  $d>5$ the massless limit
is completely well-defined, i.e.,
it is obtained smoothly from the
massless limit of the scalar-type perturbations of the Proca field.
For vector-type perturbations
the Maxwell field
limit is obtained smoothly for all $d$ from the
massless limit of the vector-type perturbations of the Proca field.

\end{abstract}

\maketitle

\section{Introduction}\label{sec:intro}

The anti-de Sitter spacetime \cite{Calabi:1962,Penrose:1968,hawking},
or AdS for short, is the maximally symmetric vacuum solution to the
Einstein field equations with a negative cosmological constant. This
spacetime can be obtained by performing the universal cover of the
AdS universe. As a result, the AdS spacetime is
not a globally hyperbolic spacetime and it possesses a timelike
boundary at spatial infinity. 

Due to the properties at spatial infinity, there is 
an intrinsic interest in asymptotically AdS spacetimes as they can
describe systems in a gravitational box. In particular, pure
AdS spacetime is of special importance in the construction of quantum
field theories \cite{avis}, where
two reflective and one transparent
boundary conditions to the fields are possible at infinity.
AdS spacetime is also essential in the formulation of supergravity
theories, where it acts as a natural arena
for quantum supersymmetric fields, including the possibility
that these might have
negative mass \cite{Breitenlohner:1982}.
In this connection, AdS 
plays a fundamental role
in the AdS/CFT conjecture, that establishes a duality
between supergravity as a low-energy
phenomenon of string theory in AdS, and a conformal field theory
at its boundary \cite{maldacena}.

Given the importance of the AdS spacetime it is
relevant to study and understand
its stability.
To linear 
perturbations AdS is stable, as we will see below, but to nonlinear 
perturbations it seems that AdS is unstable. This 
instability was explored in \cite{bizon1}, where the evolution of 
the Einstein-Klein-Gordon system was considered
and a large class of arbitrary small amplitudes of 
the scalar field was found to have
an evolution leading 
to black hole formation. It was further 
reinforced by analyzing a complex scalar field \cite{Buchel:2012}, 
yet it does not occur for all classes of 
initial data as there are islands of stability \cite{Masachs:2019}.

Linear stability of a spacetime is also important.  To study it, one
must analyze linear perturbations which are described by normal modes
for the case of pure AdS and quasinormal modes in black hole
spacetimes be they asymptotically flat, AdS, or otherwise.  In
spherical symmetry, the perturbations are classified as scalar-type,
vector-type, and tensor-type perturbations, which regards their
tensorial behavior on the 2-sphere. This decomposition allows to write
the linearized field equations as a radial Schr\"odinger-like equation
with an effective given potential for each type of perturbation.
Moreover, to obtain normal modes and quasinormal modes, one must
impose boundary conditions at the center or at the black hole horizon
if there is one, and at spatial infinity.  The linear stability of a
Schwarzschild black hole spacetime was solved first trough the
Regge-Wheeler formalism by carefully expanding the perturbations
\cite{regge_wheeler,zerilli}.  In \cite{ckjpsl} boundary conditions
and all types of perturbations in the Schwarzschild-AdS black hole
were imposed and worked out.  In \cite{konoplya} it was analyzed Proca
massive vector field perturbations in a Schwarzschild background.  In
\cite{redkov} spherical waves of spin-1 particle in anti-de Sitter
spacetime were performed. In \cite{dolan} massive vector fields on
the Schwarzschild spacetime were further analyzed.  For other boundary
conditions based on the vanishing of the energy flux see
\cite{Herdeiro2}.
In \cite{fhlc} quasinormal modes of Proca fields in a
Schwarzschild-AdS spacetime were found.  In \cite{tf} the normal modes
of Proca fields in AdS spacetime were found.

Higher dimensions are important. The universe may
be higher dimensional somewhere,
higher dimensions may have existed at some time
in the very early universe, or
perhaps they can be constructed or detected
in a future experiment.  The study of the
physics in higher dimensions
also provides a means of understanding
what is intrinsic and important to $d=4$.
AdS spacetime can be
extended to higher dimensions which is referred
to as 
AdS$_d$, where $d$ is the number of dimensions.
Normal modes and
quasinormal modes have also been studied in AdS$_d$
spacetimes.
In spherical symmetry in higher
dimensions, the
perturbations are also classified as scalar-type, vector-type, and
tensor-type perturbations, which regards their behavior now on
the $(d-2)$-sphere. Again,
the decomposition allows us to write the linearized
field equations as a radial Schr\"odinger-like equation for each
type of perturbation, whose potential now also depends on the
dimension of the spacetime.
The
normal modes of AdS$_d$  were
first obtained in \cite{burgess} for a scalar field.  The problem of
linear stability in higher-dimensional spacetimes was considered in
the Ishibashi-Kodama formalism
by expanding
the perturbations in
higher-dimensional scalar, vector, and tensor spherical harmonics,
which independently form a complete basis on a $(d-2)$-sphere
\cite{ishibashi_kodama1}.  In
\cite{ishibashiwald} it was found
that in some cases the Dirichlet boundary
conditions are not the only suitable boundary conditions and
generalized Robin boundary conditions are also permitted,  a
general analysis of the equations that the scalar, electromagnetic and
gravitational perturbations obey in AdS$_d$ was made, and a general
formula for the eigenfrequencies in a range of parameters of the
equation was obtained.  Moreover, scalar-type, vector-type and
tensor-type gravitational perturbations were studied and the
eigenfrequencies given  imposing Dirichlet boundary
conditions \cite{natario}.
The result was extended to the scalar-type and
vector-type Maxwell electromagnetic perturbations in \cite{ortega}.  The
problem of linear stability in higher-dimensional spacetimes was
further addressed by expanding
perturbations in higher-dimensional scalar, vector and tensor
spherical harmonics, which independently form a complete basis on a
$(d-2)$-sphere, now with applications
to black hole spacetimes
\cite{ishibashi_kodama2}.  In
\cite{Herdeiro} the wave equation for a Proca field in $d$-dimensional
spherically symmetric black hole spacetimes was obtained
with interest in understanding the Hawking
radiation for a Proca field in $d$-dimensions.  In \cite{ueda} massive
vector field perturbations on extremal and near-extremal static black
holes were analyzed.  
A study of the Proca perturbations in $d$-dimensional pure AdS is thus of
interest.  The mass of the Proca field introduces a coupling between
two scalar-type degrees of freedom. For black hole spacetimes in
general, a decoupling of these two degrees of freedom does
not seem to be analytically allowed. For pure AdS in four dimensions,
however, the scalar-type degrees of freedom can be decoupled by making
a transformation of the fields, and it is thus of interest to
know if this occurs in AdS$_d$.

%
%
In this work we obtain the exact expression for normal modes of linear
Proca perturbations in AdS$_d$ background, using the Ishibashi-Kodama
formalism.  We find that the scalar-type degrees of freedom decouple
in AdS$_d$, by making a linear transformation to the
relevant fields.  We also study the electromagnetic perturbations in
order to understand the $\mu = 0$ limit of the Proca field.  We
consulted results in \cite{lovenumbers,myers,higuchi} on
$(d-2)$-sphere,  and use results of the priceless
manual \cite{abramowitz}.

The work is organized as follows.
In Sec.~\ref{sec:equations} we introduce the field equations for a
Proca field minimally coupled to curved spacetime background.
In Sec.~\ref{sec:lp}, we obtain the equations for the Proca field
perturbations in pure AdS spacetime by introducing the
Ishibashi-Kodama formalism and further decomposing the Proca field in
scalar-type and vector-type components, according to their tensorial
behavior on the $(d-2)$-sphere.  We also decouple
the scalar-type components by making a linear transformation to the
relevant fields.
In Sec.~\ref{sec:lpadsd2}, we obtain the normal mode
eigenfrequencies and eigenfunctions as a solution to the Proca field
equations, which can be put into Schr\"odinger-like equations.  In
Sec.~\ref{sec:maxwell}, we rederive the normal mode frequencies in
AdS$_d$ of the Maxwell
electromagnetic perturbations and analyze it in the
context of the zero mass limit of the Proca field.
In Sec.~\ref{sec:concl}, we conclude. 
In Appendix~\ref{sec:app1}, we review the properties of spherical
harmonics on the $(d-2)$-sphere.
In Appendix \ref{sec:app2}, we study the solutions of the
hypergeometric differential equation.

\section{Proca field in curved spacetime}
\label{sec:equations}

The action of a Proca field, i.e., a massive vector field, minimally
coupled to the metric
field of a generic $d$-dimensional curved spacetime
with negative cosmological constant can be written as
\begin{align}\label{eq:actiontotal}
    S = S_{\rm EH} + S_{\rm P}\,\,,
\end{align}
where
\begin{align}\label{eq:einsteinhilbertaction}
S_{\rm EH} = \int d^{d}x \sqrt{-g}\,\frac{R -
2\Lambda}{16\pi}\,\,,
\end{align}
is the Einstein-Hilbert action,
\begin{align}\label{eq:procaaction}
S_{\rm P} = -\int d^{d}x \sqrt{-g}\left(\frac{1}{2} \mu^{2}
A_{\mu}A^{\mu} +\frac{1}{4}F_{\mu \nu}F^{\mu \nu} \right)\,\,,
\end{align}
is the Proca action, $g$ is the determinant of the metric $g_{\mu
\nu}$, ${R} = R_{\mu \nu}g^{\mu \nu}$ is the Ricci scalar defined as
the trace of the Ricci tensor $R_{ab}$ composed by the metric itself
and its first and second derivatives, $\Lambda = -
\frac{(d-1)(d-2)}{2l^2}$ is the cosmological constant with $l$ being
the characteristic AdS length, $A_\mu$ is the Proca field with mass
$\mu$ and $F_{\mu \nu} \equiv \nabla_{\mu}A_\nu-\nabla_{\nu}A_\mu$ is
the Proca field strength.
Spacetime indices are denoted by Greek letters, e.g.,
$\mu$, $\nu$, run from 0 to $d-1$, where $0$ is the time
index, and 1 to $d-1$ specify the spatial indices.
The field equations for the metric and the Proca field are obtained by
applying the variational principle to the action given 
Eq.~\eqref{eq:actiontotal}.  The field equations for the metric are
then the Einstein equations given by
\begin{equation}
\label{eq:efe}
G_{\mu \nu} - \frac{(d-1)(d-2)}{2l^2} g_{\mu \nu} =
8\pi T_{\mu \nu}\,,
\end{equation}
where $G_{\mu \nu}= {R}_{\mu \nu}-\frac{1}{2}g_{\mu \nu}{R}$ is the
Einstein tensor and $T_{\mu \nu}$ is the Proca stress-energy tensor,
given by
\begin{align}
\label{eq:energytensor}
T_{\mu \nu} &= g^{\alpha \beta}F_{\mu \alpha}F_{\nu \beta} 
 + \mu^2 A_{\mu}A_{\nu}\nonumber \\
&-g_{\mu \nu}\left(\frac{1}{4}F_{\alpha \beta}F^{\alpha \beta}
+\frac{\mu^2}{2}A_{\alpha}A^{\alpha}\right)\,.
\end{align}
Moreover, the Einstein tensor obeys the Bianchi identities 
$\nabla^\mu G_{\mu \nu} =0$, which in turn imply the conservation law for 
$T_{\mu \nu}$, i.e., $\nabla^\mu T_{\mu \nu} = 0$. The Proca field equations 
are obtained either by the conservation law for $T_{\mu \nu}$ or by varying 
the action with respect to the Proca field $A_\mu$ and can be written as
\begin{equation}\label{eq:proca_equation}
    \nabla_\nu F^{\mu \nu}+ \mu^2A^\mu = 0\,.
\end{equation}
Due to $F_{\mu \nu}$ being an antisymmetric tensor, one can calculate
the divergence of Eq.~\eqref{eq:proca_equation} to obtain a Bianchi
identity for $A_\mu$
\begin{equation}
\label{eq:bianchi}
\nabla^{\mu}A_{\mu}= 0\,.
\end{equation}
It must be noted that Eq.~\eqref{eq:bianchi} is a direct consequence
of the Proca
field equation, Eq.~\eqref{eq:proca_equation}, when $\mu \ne 0$.
Thus, $A_\mu$ is a physical field and describes $d-1$ degrees of
freedom, as one component of the vector field can always be obtained
from the others by integrating Eq.~\eqref{eq:bianchi}. For $\mu = 0$,
$A_\mu$ corresponds to the Maxwell field and the field equation
Eq.~\eqref{eq:proca_equation}, becomes invariant under the gauge
transformation
$\label{eq:gauge_transf}
A^\mu \longrightarrow A^\mu+\partial^\mu h
$,
where $h$ is an arbitrary scalar field. The Bianchi identity
for Proca fields,
Eq.~\eqref{eq:bianchi}, ceases
to be a consequence of the field equation and becomes the usual
Lorenz gauge condition of the Maxwell field. Even after imposing this
gauge, a residual gauge freedom remains as Eq.~\eqref{eq:bianchi} is
invariant under $A^\mu \longrightarrow A^\mu+\partial^\mu h$
if $\nabla_\mu \nabla^\mu
h = 0$, i.e., if $h$ obeys the Klein-Gordon equation. Hence, the
Maxwell field describes $d-2$ degrees of freedom.  In $d=4$, the
previous discussion implies that while the Proca field describes three
degrees of freedom, corresponding to two transversal polarizations and
one longitudinal polarization, the residual gauge freedom of the
Maxwell field eliminates the longitudinal polarization, so that in
total the Maxwell field describes two degrees of freedom,
corresponding to the two transversal polarizations.

To formally describe the spacetime permeated by the Proca field, one
would have to solve Eq.~\eqref{eq:efe} for the metric $g_{\mu \nu}$
and Eq.~\eqref{eq:proca_equation} for the Proca field $A_\mu$
simultaneously.  Here we are only interested in linear perturbations
of the Proca field $A_{\mu}$ around the trivial solution
$A_{\mu}=0$. As a result, we only consider Eqs.~\eqref{eq:efe}
and~\eqref{eq:proca_equation} up to first order in perturbations of
the Proca field.  Since $T_{\mu \nu}$ in Eq.~\eqref{eq:energytensor}
is of second order in $A_\mu$, linear perturbations in the Proca field
induce a curvature perturbation on $g_{\mu \nu}$ only at second order
and $T_{\mu \nu}$ can be neglected in Eq.~\eqref{eq:efe} at first
order, reducing Eq.~\eqref{eq:efe} to the vacuum Einstein equations
with negative cosmological constant.  Thus, $g_{\mu \nu}$ corresponds
to the background metric, as if the Proca field $A_\mu$ was
absent. Moreover, the linear Proca perturbations obey the Proca
equation, Eq.~\eqref{eq:proca_equation}, with
the connection $\Gamma^\mu_{\nu\sigma}$ being
associated to the background metric.

In what follows, we apply the
Ishibashi-Kodama formalism 
to the Proca equation
in a $d$-dimensional AdS spacetime. We firstly write the Proca
equation and decompose the $A_\mu$ field in the
AdS\textsubscript{$d$} background.

\section{Linear Proca field perturbations in AdS$_d$}
\label{sec:lp}

\subsection{The factorization of AdS{$_d$} spacetime
and the Proca equation}

We consider the $d$-dimensional AdS spacetime
which is a solution of the vacuum Einstein equation
with a cosmological constant, Eq.~\eqref{eq:efe}. 
The $d$-dimensional AdS spacetime manifold, $\mathcal{M}^d
\equiv {\rm AdS}_d$, 
$d \geq 4$,
can be written 
as a warped product of a submanifold $\mathcal{N}^2$
of dimension two with a $(d-2)$-sphere,  $\mathcal{S}^{d-2}$,
i.e., 
$\mathcal{M}^d=\mathcal{N}^2 \times \mathcal{S}^{d-2}$.
Its associated line element $g_{\mu\nu}dx^\mu dx^\nu$
in coordinates $x^\mu$, $\mu=0,...,d-1$, is thus decomposed
as
\begin{align}
g_{\mu\nu}dx^\mu dx^\nu = 
{\tilde g}_{ab}dy^a dy^b+
r^2{\hat g}_{ij}d\theta^i d\theta^j \,,
\label{eq:adsmetric}
\end{align}
where 
\begin{align}
&{\tilde g}_{ab}dy^a dy^b =  - f(r) dt^2 + \dfrac{dr^2}{f(r)}
\,,\quad f(r) = 1 + \frac{r^2}{l^2}
\,,\label{eq:hmetric}
\end{align}
is the line element of the $\mathcal{N}^2$ submanifold,
written in coordinates $y^a = (y^0,y^1)=(t,r)$, with $t$
and $r$ being the time and radial coordinates,
respectively,
and 
\begin{align}
{\hat g}_{ij}d{\theta}^i d{\theta}^j  = 
(d\theta^2)^2 + \sum_{i=3}^{d-1} \prod^{i-1}_{k=2} 
\sin^2(\theta^k) (d\theta^i)^2\,,
\label{eq:spheremetric}
\end{align}
which is the line element of the $(d-2)$-sphere, $\mathcal{S}^{d-2}$,
sometimes represented as $(d\Omega^{d-2})^2$,
where the $\theta^i$ are the angular coordinates 
and $i,j,k=2,...,d-1$.
Note that in this convention $\theta^2$ would be the usual $\theta$
and $\theta^3$ would be the usual $\phi$ in $d=4$.
As it should be clear by now,
to allow one to distinguish between tensors living on the different
manifolds $\mathcal{M}^d$, $\mathcal{N}^2$, and $\mathcal{S}^{d-2}$, we
use Greek indices $\mu,\nu,...$ for tensors on $\mathcal{M}^d$, latin
indices in the range $a,b,...,h$ for tensors on $\mathcal{N}^2$ and
latin indices in the range $i,j,...$ for tensors on
$\mathcal{S}^{d-2}$.

In order to factorize AdS$_d$ spacetime, one also must separate
the connection $\nabla_\mu$ associated to the manifold
$(\mathcal{M}^d, g)$ into the connection $\tilde{\nabla}_a$ associated
to the manifold $(\mathcal{N}^2,\tilde{g})$ and the connection
$\hat{\nabla}_i$ associated to the manifold
$(\mathcal{S}^{d-2},\hat{g})$.  This can be accomplished by using the
following relations between the Christoffel symbols associated to each
manifold, those are
\begin{align}\label{eq:chris}
&\Gamma^a_{\ b c} = \tilde{\Gamma}^a_{\ b c} \,, \quad 
\Gamma^a_{\ i j} = -r \left(\partial^a r \right){\hat g}_{i j}
\,,\quad \notag\\ 
&\Gamma^i_{\ a j} = \frac{\partial_a r}{r}\delta^i_j
\,, \quad 
\Gamma^i_{\ j k} = \hat{\Gamma}^i_{\ j k} \,,
\end{align}
where
${\Gamma}^\mu_{\ \nu\rho}$
are the
Christoffel symbols associated with the metric 
$g_{\mu\nu}$, with the greek indices spanning through 
the indices in $\mathcal{N}^2$ and 
in $\mathcal{S}^{d-2}$, e.g. $\mu = \{a,i\}$, and 
$\tilde{\Gamma}^a_{\ b c}$ and $\hat{\Gamma}^i_{\ j k}$ are the
Christoffel symbols associated with the metrics ${\tilde g}_{a b}$ and
${\hat g}_{i j}$, respectively.

The projections of the Proca equation, 
Eq.~\eqref{eq:proca_equation}, into $\mathcal{N}^2$ and
$\mathcal{S}^{d-2}$ are written as
\begin{equation}\label{eq:eom1}
\tilde{\nabla}_b F^{a b} + (d-2) \frac{\partial_b r}{r} F^{a b} 
+ \hat{\nabla}_j F^{a j}+ \mu^2 A^a = 0 \,,
\end{equation}
\begin{equation}\label{eq:eom2}
\tilde{\nabla}_b F^{i b} + (d-2) \frac{\partial_b r}{r} F^{i b} 
+ \hat{\nabla}_j F^{i j}+ \mu^2 A^i = 0\,,
\end{equation}
respectively.
These equations are supplemented with the Bianchi
identity, Eq.~\eqref{eq:bianchi}, which is now
\begin{equation}\label{eq:eom3}
\tilde{\nabla}_a A^a + (d-2) \frac{\partial_a r}{r}A^a+
\hat{\nabla}_i A^i = 0\,.
\end{equation}

\subsection{The decomposition of the Proca field and spherical
harmonics expansion}

To simplify the field equations for $A_\mu$, one exploits the
spherical symmetry of the background metric in
Eq.~\eqref{eq:adsmetric}. The strategy is to project the field $A_\mu$
into components that are orthogonal to $\mathcal{S}^{d-2}$ and
components that are tangent to $\mathcal{S}^{d-2}$. The $A_\mu$ field
can be written as
\begin{equation}\label{eq:gauge1}
A_\mu dx^\mu = {\tilde A}_a dy^a +
{\hat A}_i d{\theta}^i \,,
\end{equation}
where ${\tilde A}_a$ denotes the projection of $A_\mu$
orthogonal to the cotangent space of $\mathcal{S}^{d-2}$ and
${\hat A}_i$ denotes the projection of $A_\mu$ tangent to the
cotangent space of $\mathcal{S}^{d-2}$.  The latter can be further
decomposed using the Helmoltz-Hodge decomposition
\cite{ishibashi_kodama1}, which allows one to write uniquely a dual
vector field on $\mathcal{S}^{d-2}$, ${\hat A}_i$, as the sum
of a scalar field on $\mathcal{S}^{d-2}$, ${\hat A}^{(s)}$,
and a transverse
covector field on the cotangent space of $\mathcal{S}^{d-2}$,
${\hat A}^{(v)}_i$, in the following way
\begin{equation}\label{eq:gauge2}
{\hat A}_i = {\hat A}^{(v)}_i +\hat{\nabla}_i {\hat A}^{(s)}
\quad \,, \quad \quad 
\hat{\nabla}_i {{\hat A}^{(v)\,i}} = 0\,.
\end{equation}
Since ${\tilde A}_a$ and ${\hat A}^{(s)}$ behave as scalars on
$\mathcal{S}^{d-2}$, these are called the scalar-type components of
$A_\mu$, while ${\hat A}^{(v)\,i} = {\hat g}^{ij} {\hat A}^{(v)}_j$,
a vector on the
tangent space of $\mathcal{S}^{d-2}$, is called the vector-type
component of $A_\mu$.
Moreover, the scalar-type
components of the Proca field transform as scalars  under the
SO($d-1$) rotation group, and
the vector-type
component of the Proca field transforms as
a vectors under the
SO($d-1$) rotation group.
Since the correspondence between ${\hat A}^{(v)\,i}$ and ${\hat
A}^{(v)}_i$ is one to one, we shall make the abuse of language that
${\hat A}^{(v)}_i$ corresponds to the vector-type component of $A_\mu$
as well.  In the literature, scalar-type is also referred to as polar-
or even-type, whereas vector-type is called axial- or odd-type.
This has to do with the transformation properties of these components
under parity transformations, see \cite{regge_wheeler, zerilli}. The
scalar-type components of $A_\mu$ can be expanded in scalar harmonics
$Y_{\vec{k}_s}$, where $\vec{k}_s$ is a vector containing the the
angular momentum number $\ell$ and the $d-3$ azimuthal numbers, which
form a complete basis on $\mathcal{S}^{d-2}$, satisfying
\begin{align}
\label{eq:scalar_harmonics}
\left(\hat{\Box}+k_{s}^2\right) Y_{\vec{k}_s} = 0 \,, \quad\quad
\int d{\Omega}^{d-2} Y_{\vec{k}_s} Y_{\vec{k}_s'} =
\delta_{\vec{k}_s \vec{k}_s'} \,,
\end{align}
where $ \hat{\Box} = {\hat g}^{i j}\hat{\nabla}_i\hat{\nabla}_j $,
\begin{equation}\label{eq:eigenvalues_scalar}
k_s^2 = \ell(\ell+d-3) \,, \quad \ell = 0,1,2,... \,,
\end{equation}
and $d{\Omega}^{d-2}$ is here
the volume element over the sphere
given by $d{\Omega}^{d-2}=\sqrt{\hat g}d\theta^2...
d\theta^{d-1}$, $\hat g$ being the determinant of
the metric ${\hat g}_{ij}$.
Similarly, the
vector-type component of $A_\mu$ can be expanded in vector harmonics
${Y_{\vec{k}_v\,i}}$ which also form a complete basis on
$\mathcal{S}^{d-2}$, where $\vec{k}_v$ is a vector containing the
angular momentum number $\ell$ and $d-3$ azimuthal numbers, that in
principle are different from the azimuthal numbers of the scalar
harmonics. The vector harmonics ${Y_{\vec{k}_v\,i}}$ then satisfy
\begin{align}\label{eq:vector_harmonics}
&\left(\hat{\Box}+k_{v}^2\right) {Y_{\vec{k}_v\,i}} = 0 \,, \quad
\hat{\nabla}_i {Y_{\vec{k}_v}^i} = 0\,\,,\nonumber\\
&\int d{\Omega}^{d-2}
{\hat g}^{i j}{Y_{\vec{k}_v\,i}} {Y_{\vec{k}_v'\,j}} 
= \delta_{\vec{k}_v \vec{k}_v'} \,,
\end{align}
with
\begin{equation}\label{eq:eigenvalues_vector}
k_v^2 = \ell(\ell+d-3)-1 \,, \quad \ell = 1,2,3,...\,,
\end{equation}
see Appendix \ref{sec:app1} for more details.

Thus, 
the scalar-type components $\{{\tilde A}_a,{\hat A}^{(s)}\}$
can be expanded
in terms of  the
$Y_{\vec{k}_s}(\theta)$, and 
vector-type component $\{{\hat A}_i^{(v)}\}$
can be expanded
in terms of  the
${Y_{\vec{k}_v\,i}}(\theta)$,
where $\theta\equiv(\theta^2...,\theta^i,...\theta^{d-1})$.
Indeed, $\{{\tilde A}_a,{\hat A}^{(s)}\}$
can be expanded as 
${\tilde A}_a (y,{\theta}) = \sum_{\vec{k}_s}
{{\tilde A}_{\vec{k}_s\,a}}(y) Y_{\vec{k}_s}({\theta})$,
where $y\equiv(t,r)$,
and  ${\hat A}^{(s)} (y,{\theta}) =
\sum_{\vec{k}_s} {{\hat A}^{(s)}_{\vec{k}_s}}
(y) Y_{\vec{k}_s}(\theta)$, respectively,
 and 
$\{{\hat A}_i^{(v)}\}$ as
${\hat A}^{(v)}_i (y,{\theta}) = \sum_{\vec{k}_v} 
 {\hat A}^{(v)}_{\vec{k}_v}(y) {Y_{\vec{k}_v\,i}}({\theta})$.
Note that
$\hat{\nabla}_i {\hat A}^{(s)} (y,{\theta}) =
\sum_{\vec{k}_s} {{\hat A}^{(s)}_{\vec{k}_s}}
(y) \hat{\nabla}_i Y_{\vec{k}_s}(\theta)$.
Since ${\tilde A}_{\vec{k}_s\,a}$,
${\hat A}^{(s)}_{\vec{k}_s}$,
and 
${\hat A}^{(v)}_{\vec{k}_v}$
are cumbersome symbols to
carry along, we define
$\psi_{\vec{k}_s\,a}\equiv {\tilde A}_{\vec{k}_s\,a}$,
$\phi_{\vec{k}_s}\equiv {\hat A}^{(s)}_{\vec{k}_s}$,
and
$\chi_{\vec{k}_v}\equiv {\hat A}^{(v)}_{\vec{k}_v}$.
so
that for
the scalar-type components $\{{\tilde A}_a,{\hat A}^{(s)}\}$,
one writes
\begin{equation}\label{eq:scalar_type_vector}
{\tilde A}_a (y,{\theta}) = \sum_{\vec{k}_s}
{\psi_{\vec{k}_s\,a}}(y) Y_{\vec{k}_s}({\theta})\,,
\end{equation}
\begin{equation}\label{eq:scalar_type_scalar}
{\hat A}^{(s)} (y,{\theta}) =
\sum_{\vec{k}_s} {\phi_{\vec{k}_s}}
(y) Y_{\vec{k}_s}(\theta) \,,
\end{equation}
while for the vector-type component $\{{\hat A}_i\}$ one writes
\begin{equation}\label{eq:vector_type_scalar}
{\hat A}^{(v)}_i (y,{\theta}) = \sum_{\vec{k}_v} 
\chi_{\vec{k}_v}(y) {Y_{\vec{k}_v\,i}}({\theta})\,.
\end{equation}
Here,
${\psi_{\vec{k}_s\,a}} (y)$ is a vector field on
$\mathcal{N}^2$, and 
$\phi_{\vec{k}_s} (y)$ and $\chi_{\vec{k}_v}(y)$ are scalar
fields on $\mathcal{N}^2$. 
The scalar-type components cover two degrees of freedom, whereas the
vector-type component covers $d-3$ degrees of freedom
as can be noted from the
transverse condition of ${Y_{\vec{k}_v\,i}}$ in
Eq.~\eqref{eq:vector_harmonics}. As Eq.~\eqref{eq:bianchi} needs to
be satisfied, these variables cover in total $d-1$ degrees of freedom,
as expected for a Proca field. In summary, the complete expression of
the Proca field decomposed in spherical harmonics is
\begin{align}
A_{\mu}dx^{\mu}
\hskip-0.09cm
=
\hskip-0.09cm
\sum_{\vec{k}_s}
\hskip-0.09cm
\left[ {\psi_{\vec{k}_s\,a}} Y_{\vec{k}_s} dy^a
\hskip-0.09cm
+
\hskip-0.09cm
\phi_{\vec{k}_s} \hat{\nabla}_i Y_{\vec{k}_s} d\theta^i\right]
\hskip-0.09cm
+
\hskip-0.09cm
\sum_{\vec{k}_v} \chi_{\vec{k}_v} Y_{\vec{k}_v\,i} d\theta^i .
\label{eq:Afieldcomplete}
\end{align}

In terms of the expansion in
Eqs.~\eqref{eq:scalar_type_vector}-\eqref{eq:vector_type_scalar},
i.e., 
Eq.~\eqref{eq:Afieldcomplete},
the components of the
Proca field strength tensor are written as
\begin{equation}\label{eq:expansion1}
F^{a b} = \sum_{\vec{k}_s} 
\left[{\tilde{\nabla}}^a \psi_{\vec{k}_s}^b 
- {\tilde{\nabla}}^b \psi_{\vec{k}_s}^a \right]
 {Y}_{\vec{k}_s} \,,
\end{equation}
\begin{align}
F^{a i} = &\sum_{\vec{k}_s} 
\left[\frac{1}{r^2}\left(  \nabla^a \phi_{\vec{k}_s}
- \psi_{\vec{k}_s}^a \right)\right] 
\hat{\nabla}^i {Y}_{\vec{k}_s} \notag\\ +&\sum_{\vec{k}_v} 
\left[\frac{1}{r^2}\nabla^a \chi_{\vec{k}_v} \right]
{Y_{\vec{k}_v}^i} ,
\label{eq:expansion2}
\end{align}
\begin{equation}\label{eq:expansion3}
F^{i j} = \sum_{\vec{k}_v} \left[\frac{\chi_{\vec{k}_v}}{r^4} \right]
\left(\hat{\nabla}^i Y_{\vec{k}_v}^j- \hat{\nabla}^j
Y_{\vec{k}_v}^i\right)\,,
\end{equation}
where the covariant derivative $\nabla_a$
can be swapped to a partial derivative $\partial_a$
when acting on scalars.

With these decompositions, it is possible to 
decouple only partially the
equations in general.
We will see that the
scalar-type perturbations are completely decoupled
from the vector-type perturbations. Moreover,
the separation of the Proca equations are achieved due to the
separation of the fields into functions that purely depend on the
coordinates $(t,r)$ and the spherical harmonics that depend only on
the spherical angles.

\subsection{Separation of the Proca equations}

\subsubsection{Proca equations after spherical harmonics
expansion}

The expansion on spherical harmonics of the Proca field in
Eqs.~\eqref{eq:scalar_type_vector}-\eqref{eq:vector_type_scalar},
i.e., 
Eq.~\eqref{eq:Afieldcomplete},
and the
corresponding expansion of the strength field tensor in
Eqs.~\eqref{eq:expansion1}-\eqref{eq:expansion3} can be inserted into
Eqs.~\eqref{eq:eom1}-\eqref{eq:eom3}, which allows the separation of
the Proca equations in AdS$_d$. In
this way, the Proca equations are
separated into three sums, one in terms of the spherical harmonics
$Y_{\vec{k}_s}$ over each $\vec{k}_s$, i.e.,
$\sum_{\vec{k}_s}{\tilde W}_{\vec{k}_s}^a Y_{\vec{k}_s}$, another in
terms of the gradients of the spherical harmonics $\hat{\nabla}_i
Y_{\vec{k}_s}$ over each $\vec{k}_s$,
i.e., $\sum_{\vec{k}_s}{\hat W}_{\vec{k}_s}^{(s)} \hat{\nabla}^i
Y_{\vec{k}_s}$, and the remaining sum in terms of the vector spherical
harmonics $Y_{\vec{k}_v\,i}$ over each $\vec{k}_v$, i.e., 
$\sum_{\vec{k}_v}{\hat W}_{\vec{k}_v}^{(v)} Y_{\vec{k}_v}^i$, where the
coefficients of all the three sums only depend on the 
coordinates $t$ and $r$, i.e., ${\tilde W}_{\vec{k}_s}^a =
{\tilde W}_{\vec{k}_s}^a(t,r)$, 
${\hat W}_{\vec{k}_s}^{(s)} = {\hat W}_{\vec{k}_s}^{(s)} (t,r)$,
and ${\hat W}_{\vec{k}_v}^{(v)} = {\hat W}_{\vec{k}_v}^{(v)}(t,r)$.
We must note
that $Y_{\vec{k}_s}$, $\hat{\nabla}_i Y_{\vec{k}_s}$ and
$Y_{\vec{k}_v\,i}$ are orthogonal between each other.  An argument can
be made that these span different representations of the rotation
group $SO(d-1)$, for $d>4$, while for $d=4$, the scalar and vector
spherical harmonics span different representations of the rotation
group $O(3)$, see \cite{ishibashiwald} and also Appendix~\ref{sec:app1}.
Therefore, the Proca equations separate into the equations
$\tilde{W}_{\vec{k}_s}^a = 0$, $W_{\vec{k}_s}^{(s)}=0$, and
$W_{\vec{k}_v}^{(v)} = 0$, described only by the coordinates $t$ and
$r$.  Since the Proca equation is a linear differential equation in
$A_\mu$, there is no mixing between the different $\vec{k}_v$ and
$\vec{k}_s$ modes. Therefore, without loss of generality and for
convenience, we drop the sum on all the $\vec{k}_s$ and
$\vec{k}_v$. In the rest of the section, we treat the scalar spherical
harmonics as $Y$ and the vector spherical harmonics as $Y_i$.

We now show the equations obtained from the separation of the Proca
equations.  The projection of the Proca equation into $\mathcal{N}^2$,
Eq.~\eqref{eq:eom1}, can be written in terms of a sum of spherical
harmonics $Y$, in which the associated coefficients must satisfy
${\tilde W}_{\vec{k}_s}^a = 0$, or explicitly,
\begin{align}
& 2 \tilde{\nabla}_b \tilde{\nabla}^{[a} \psi^{b]} 
+ 2 (d-2) \frac{\tilde{\nabla}_b r}{r}\tilde{\nabla}^{[a} \psi^{b]} 
\notag \\ &+ \left(\frac{\ell(\ell+d-3)}{r^2}+\mu^2\right)\psi^a
-\frac{\ell(\ell+d-3)}{r^2}\partial^a \phi = 0 \,,
\label{eq:eomScalar1}
\end{align}
for each $\vec{k}_s$, where it was used that $k^2_s=\ell(\ell+d-3)$,
$\tilde{\nabla}^{[a} \psi^{b]} = \frac{1}{2}(\tilde{\nabla}^{a}
\psi^{b} - \tilde{\nabla}^{b} \psi^{a})$, and that
$\hat{\nabla}_i Y^i = 0$, which avoids the appearance of $\chi$ in
Eq.~\eqref{eq:eomScalar1}.
The projection of the Proca equation into $\mathcal{S}^{d-2}$,
Eq.~\eqref{eq:eom2}, can be written in terms of two sums.
One sum is in terms
of the gradient of the spherical harmonics $\hat{\nabla}_i Y$, whose
coefficients satisfy ${\hat W}_{\vec{k}_s}^{(s)}=0$, or explicitly
\begin{align}\label{eq:eomScalar2}
\tilde{\Box}{\phi}
\hskip-0.05cm
+
\hskip-0.05cm
(d-4) \frac{\partial_b r}{r}\partial^b \phi
\hskip-0.05cm
-
\hskip-0.05cm
\mu^2\phi
\hskip-0.05cm
-
\hskip-0.05cm
{\tilde{\nabla}}_b  \psi^b 
\hskip-0.05cm
-
\hskip-0.05cm
(d-4)\frac{\partial_b r}{r} \psi^b
\hskip-0.05cm=\hskip-0.05cm 0,
\end{align}
for each $\vec{k}_s$, where
$\tilde{\Box}=\tilde{g}^{ab}\tilde{\nabla}_a\tilde{\nabla}_b$.
The other sum is in
terms of the vector spherical harmonics $Y_i$, whose coefficients must
satisfy ${\hat W}_{\vec{k}_v}^{(v)} = 0$ or explicitly
\begin{align}
\tilde{\Box} \chi
\hskip-0.05cm
+
\hskip-0.05cm
\frac{d-4}{r} (\partial^b r) 
(\partial_b \chi)
\hskip-0.05cm
-
\hskip-0.05cm
\left(\frac{\ell(\ell+d-3)\hskip-0.05cm+\hskip-0.05cm
d\hskip-0.05cm-\hskip-0.05cm4}{r^2}
\hskip-0.05cm
+
\hskip-0.05cm
\mu^2\right)
\hskip-0.05cm
\chi
\hskip-0.05cm
=
\hskip-0.05cm
0,
\label{eq:eomVector}
\end{align}
for each $\vec{k}_v$, where $k^2_v = \ell(\ell+d-3) - 1$ was used.
Furthermore, to obtain Eq.~\eqref{eq:eomVector}, the commutator
$2\hat{\nabla}_{[j}\hat{\nabla}_{i]} Y^j = \hat{R}_{m i} Y^m$ was
used, where $\hat{R}_{m i} = (d-3) \hat{g}_{mi}$ is the Ricci tensor of
$\mathcal{S}^{d-2}$.
Finally, the Bianchi identity Eq.~\eqref{eq:eom3} can be written as a
sum of spherical harmonics $Y$, whose coefficients satisfy
\begin{align}
\tilde{\nabla}_b \psi^b + (d-2) \frac{\partial_b r}{r} \psi^b 
- \frac{\ell(\ell+d-3)}{r^2}\phi = 0\,\,,
\label{eq:eomBianchi}
\end{align}
for each $\vec{k}_s$, where again $k_s^2 =
\ell(\ell+d-3)$ and $\hat{\nabla}_i Y^i = 0$ 
were used. 
We see that Eqs.~\eqref{eq:eomScalar1}-\eqref{eq:eomVector}
form a set of four equations, two 
component equations in 
Eq.~\eqref{eq:eomScalar1} and two equations
in 
Eqs.~\eqref{eq:eomScalar2}
and \eqref{eq:eomVector}.
There is a fifth equation, the
Bianchi identity
equation for Proca fields given in Eq.~\eqref{eq:eomBianchi}.
So in total we can play with five equations.

\subsubsection{Proca equations decoupled: The important
equations for quasinormal modes}

The equations for the scalar-type components of the Proca field,
i.e., Eqs.~\eqref{eq:eomScalar1} and \eqref{eq:eomScalar2}
are coupled in $\psi_t$, $\psi_r$, 
$\phi$, and Eq.~\eqref{eq:eomVector}
for the vector-type component
represented by $\chi$ is decoupled.
The Bianchi identity Eq.~\eqref{eq:eomBianchi}
is also coupled.
By defining new variables $q_0$,
$q_1$, $q_2$, and $q_3$ as functions of $\psi_t$, $\psi_r$, 
$\phi$, and $\chi$ it is possible to decouple
the equations. We give first the result and then show
the steps to obtain it.
So,  Eqs.~\eqref{eq:eomScalar1}-\eqref{eq:eomVector}
in the new variables are 
\begin{align}
&\hat{\mathcal{D}}_{\ell} q_0 + \frac{2r}{l^2}\left(\partial_t q_1
+ \partial_t q_2-\partial_{r_*}  q_0\right) = 0 \,,
\label{eq:q0}\\
&\hat{\mathcal{D}}_{j_k} q_k = 0 \,,
\label{eq:qk}
\end{align}
and the Bianchi identity Eq.~\eqref{eq:eomBianchi}
is now
\begin{align}
&
\partial_t q_0
\hskip-0.09cm
-
\hskip-0.09cm
\partial_{r_*} (q_1
\hskip-0.09cm
+
\hskip-0.09cm
q_2)
\hskip-0.09cm
= 
\hskip-0.09cm
\frac{f}{r}
\hskip-0.09cm
\left(
\hskip-0.09cm
\frac{d
\hskip-0.09cm
-
\hskip-0.09cm
2}{2}
\hskip-0.09cm
-
\hskip-0.09cm
(\ell
\hskip-0.09cm
+
\hskip-0.09cm
d
\hskip-0.09cm
-
\hskip-0.09cm
3)
\hskip-0.09cm
\right)
\hskip-0.09cm
q_2
\hskip-0.09cm
+
\hskip-0.09cm
\frac{f}{r}
\hskip-0.09cm
\left(\ell
\hskip-0.09cm
+ \hskip-0.09cm
\frac{d
\hskip-0.09cm
-
\hskip-0.09cm
2}{2}\right)
\hskip-0.09cm
q_1 ,
\label{eq:q_bianchi}
\end{align}
where here $k\in\{1,2,3\}$,
$j_k=(j_1,j_2,j_3)$,
$j_1 = \ell +1$
with $\ell\in \mathbb{N}_0$, 
$j_2 = \ell -1$ with $\ell \in \mathbb{N}$,
and $j_3 = \ell$ with $\ell \in \mathbb{N}$,
$r_*$ is defined as
$r_* = l \arctan\left(\frac{r}{l}\right)$,
the operator $\hat{\mathcal{D}}_\ell$ is defined as 
\begin{align}
&\hat{\mathcal{D}}_\ell = -\partial_t^2 + \partial_{r_*}^2 \notag \\
&- f\left[ 
\frac{\ell(\ell+d-3)}{r^2} + \mu^2 
+ \frac{(d-2)(d-4)}{4l^2}\left(1 + \frac{l^2}{r^2}\right)\right]\,\,,
\label{eq:Doperator}
\end{align}
and 
$q_0 (t,r)$, $q_1 (t,r)$,  $q_2 (t,r)$,
and $q_3(t,r)$ are defined by
\begin{align}
q_0(t,r) =\psi_t(t,r)\,r^{\frac{d}2-1}\,,
\label{eq:q0psit}
\end{align}
\begin{align}
q_1(t,r)
\hskip-0.09cm
=
\hskip-0.09cm
\frac{(\ell
\hskip-0.09cm
-
\hskip-0.09cm
d
\hskip-0.09cm
-
\hskip-0.09cm
3)\psi_r(t,r)\,f(r)-
\ell(\ell+d-3)\frac{\phi(t,r)}{r}}{2\ell+d-3}
\,r^{\frac{d}2-1},
\label{eq:tq1psir}
\end{align}
\begin{align}
q_2(t,r) =
\frac{\ell\psi_r(t,r)\,f(r)+
\ell(\ell+d-3)\frac{\phi(t,r)}{r}}{2\ell+d-3}
\,r^{\frac{d}2-1},
\label{eq:q2phi}
\end{align}
\begin{align}
q_3(t,r) =\frac{\chi(t,r)}{r}\,r^{\frac{d}2-1}\,.
\label{eq:q3chi}
\end{align}
We see that Eqs.~\eqref{eq:q0}-\eqref{eq:qk}
provide four equations, the number we had originally,
and there is still the
Bianchi identity for Proca fields given
in Eq.~\eqref{eq:q_bianchi}, yielding
five equations in total.

\subsubsection{Proof of the decoupling of
Proca equations}

Now we show
how to obtain Eqs.~\eqref{eq:q0}-\eqref{eq:q_bianchi} together with the
definitions Eqs.~\eqref{eq:Doperator}-\eqref{eq:q3chi}.  The equations
for the scalar-type components of the Proca field, i.e.,
Eqs.~\eqref{eq:eomScalar1} and ~\eqref{eq:eomScalar2}
are coupled in $\psi_t$, $\psi_r$, and
$\phi$.
The Bianchi identity for Proca fields 
given in Eq.~\eqref{eq:eomBianchi} is also coupled in 
$\psi_t$, $\psi_r$, and
$\phi$.
We now show that
in the case of AdS$_d$ spacetime, it is
indeed possible to manipulate
these equations and decouple these components through further
transformations. We start by differentiating Eq.~\eqref{eq:eomBianchi}
to obtain
\begin{align}
\ell(\ell+d-3)\partial^a \phi = \tilde{\nabla}^a(r^2 
\tilde{\nabla}_b \psi^b) + (d-2)
\tilde{\nabla}^a(r \psi^b \partial_b r )\,.
\label{eq:Bianchideriv}
\end{align}
Using 
 Eq.~\eqref{eq:Bianchideriv}
in Eq.~\eqref{eq:eomScalar1} we obtain a coupled
equation for the $\psi^a$ components given by
\begin{align}
&\tilde{\Box} \psi^a - {\tilde{R}}^a_b \psi^b + (d-2)
\frac{\partial^b r}{r} {\tilde{\nabla}}_b \psi^a \notag \\ &-
\left(\frac{\ell(\ell+d-3)}{r^2}+\mu^2\right)\psi^a + (d-2)
{\tilde{\nabla}}^a \left(\frac{\partial_b r}{r}\right) \psi^b
\notag \\ & + \frac{2 \partial^a r
}{r}\left({\tilde{\nabla}_b}\psi^b+(d-2)\frac{\partial_b
r}{r}\psi^b\right)=0\,,
\label{eq:eomScalar1better}
\end{align}
where ${\tilde{R}}^a_b = - \frac{f''}{2}\delta^a_b$ is the Ricci
tensor of $\mathcal{N}^2$ and the commutator
$2\tilde{\nabla}_{[a}\tilde{\nabla}_{b]} \psi^a = \tilde{R}_{a b}
\psi^a$ was used.
Now, 
Eqs.~\eqref{eq:eomScalar2}
\eqref{eq:eomBianchi}, and 
\eqref{eq:eomScalar1better}
can be further
simplified into
\begin{align}
&\hat{\mathcal{D}}_{\ell} u_0 + \frac{2r}{l^2}\left(\partial_t
u_1-\partial_{r_*} u_0\right) = 0 \,,
\label{eq:u0}\\
&\hat{\mathcal{D}}_{\ell} u_1 - \frac{2f}{r^2}\left( \frac{d-2}{2}u_1
- u_2\right) = 0 \,,
\label{eq:u1}\\
&\hat{\mathcal{D}}_{\ell} u_2 +
\frac{2f}{r^2}
\left(
\left(\frac{d}{2}-2\right)
u_2+
{\ell(\ell+d-3)}u_1 \right)= 0\,,
\label{eq:u2}\\
&
\partial_t
u_0- \partial_{r_*} u_1 =
\frac{f}{r}\left(\frac{d-2}{2}u_1-u_2\right)
\,,
\label{eq:u_bianchi}
\end{align}
where again $r_*$ is defined such as
$\frac{dr_*}{dr}= \frac1{f(r)}$ with $f(r) = 1+ \frac{r^2}{l^2}$,
i.e., $r_* = l \arctan\left(\frac{r}{l}\right)$, 
$\hat{\mathcal{D}}_\ell$ is the operator
already given in Eq.~\eqref{eq:Doperator}, i.e.,
$
\hat{\mathcal{D}}_\ell = -\partial_t^2 + \partial_{r_*}^2
- f\left[ 
\frac{\ell(\ell+d-3)}{r^2} + \mu^2 
+ \frac{(d-2)(d-4)}{4l^2}\left(1 + \frac{l^2}{r^2}\right)\right]
$,
$u_0 (t,r)$, $u_1 (t,r)$, and $u_2 (t,r)$ are defined by
\begin{align}
u_0(t,r) =\psi_t(t,r)\,r^{\frac{d}2-1}
\,,\label{eq:u0psit}
\end{align}
\begin{align}
u_1(t,r) =\psi_r(t,r)\,f(r)\,r^{\frac{d}2-1}
\,,\label{eq:u1psir}
\end{align}
\begin{align}
u_2(t,r) =\frac{\phi(t,r)}{r}\,{\ell(\ell+d-3)}\,r^{\frac{d}2-1}
\,,\label{eq:u2phi}
\end{align}
and we also have used Eq.~\eqref{eq:u_bianchi}
in the last two terms of Eq.~\eqref{eq:u1}. Finally, one can notice
that the coupling terms in Eqs.~\eqref{eq:u1} and~\eqref{eq:u2} are
constants multiplied by $\frac{2f}{r^2}$. Therefore, it is possible to
further decouple Eqs.~\eqref{eq:u1} and~\eqref{eq:u2} by making the
transformations
\begin{align}
u_0 = q_0\,,
\label{eq:u0t}
\end{align}
\begin{align}
u_1 = q_1 +q_2\,,
\label{eq:u1t}
\end{align}
\begin{align}
u_2 =
(\ell+d-3)q_2 - \ell q_1\,.
\label{eq:u2t}
\end{align}
Inserting Eqs.~\eqref{eq:u0t}-\eqref{eq:u2t}
into Eqs.~\eqref{eq:u0}-\eqref{eq:u2}
yield Eqs.~\eqref{eq:q0}-\eqref{eq:qk}
with $k=1,2$, and
inserting Eqs.~\eqref{eq:u0t}-\eqref{eq:u2t}
into Eq.~\eqref{eq:u_bianchi}
yields 
Eq.~\eqref{eq:q_bianchi}.
The vector-type component of the Proca field is
described by Eq.~\eqref{eq:eomVector} and it is completely decoupled
from the scalar-type components of the Proca
field. Equation~\eqref{eq:eomVector} can be further simplified into
\begin{align}
\hat{\mathcal{D}}_\ell u_3 = 0\,\,,\label{eq:q3}
\end{align}
where
\begin{align}
u_3 = \frac{\chi}{r} \,r^{\frac{d}{2}-1}\,,
\label{eq:u3chi}
\end{align}
Defining trivially 
\begin{align}
u_3 = q_3\,.
\label{eq:u3t}
\end{align}
yields Eq.~\eqref{eq:qk} with $k=3$.
Thus, 
 Eqs.~\eqref{eq:q0}-\eqref{eq:q_bianchi} together with the
definitions Eqs.~\eqref{eq:Doperator}-\eqref{eq:q3chi}
have been obtained.

We confirm the results 
obtained in  \cite{ueda,Herdeiro}, for the particular 
case of pure AdS, where 
the vector-type component of the Proca field yields 
$d-3$ degrees of freedom, 
whereas the scalar-type component describes two 
degrees of freedom, 
which are coupled by the mass of the field. 
We find that, similarly to what was found 
in \cite{tf} for AdS$_4$, 
the scalar-type degrees of freedom decouple 
in AdS$_d$, by making a particular linear 
transformation to the relevant fields.

\section{Normal modes of Proca perturbations in the AdS$_d$ background}
\label{sec:lpadsd2}

\subsection{Initial considerations and boundary conditions}

The normal modes in AdS$_d$ are dynamical solutions of the Proca
equations described by $q_0$, $q_1$, $q_2$, and $q_3$ in
Eqs.~\eqref{eq:q0}-\eqref{eq:q_bianchi}. Moreover, the Bianchi
identity in Eq.~\eqref{eq:q_bianchi} can be used to describe $q_0$ in
terms of $q_1$, $q_2$, and $q_3$, which
in turn obey Eq.~\eqref{eq:qk}. With
the relation given by Eq.~\eqref{eq:q_bianchi}, the equation for
$q_0$, Eq.~\eqref{eq:q0}, is satisfied and so the picture is
consistent. The normal modes can then be obtained solely by solving
Eq.~\eqref{eq:qk} for the $q_1$, $q_2$, and $q_3$, with the appropriate
boundary conditions at the origin and at spatial infinity. By making
an extension to the complex numbers, one can assume  for the
$q_k (t,r)$
an ansatz of the form 
$q_k (t,r)= q_k(r) e^{-i\omega_k t}$,
where $\omega_k$ is the normal mode 
frequency of the mode
$q_k$, which is analogous to performing
a Fourier transformation from the time domain to the frequency
domain. Of course, when one wants to treat the real field, one must
project the complex field into the real axis. The equations
given in Eq.~\eqref{eq:qk}
can then be written
for the $q_k(r)$ as
\begin{align}
&\partial_{r_*}^2 q_k + \left(\omega^2_k -
V_{j_k}\right)q_k = 0\,\,,
\label{eq:eqdifnormalmodes}\\\textit{}
&V_{j_k} = f\Bigg[ 
\frac{4j_k(j_k+d-3) + (d-2)(d-4)}{4r^2} + \mu^2 
\notag\\
& \hskip 12.5em + \frac{(d-2)(d-4)}{4l^2}\Bigg]\,\,,
\end{align}
for $k\in\{1,2,3\}$.
Equation~\eqref{eq:eqdifnormalmodes} is a Schr\"{o}dinger-like
equation for the normal modes, $q_k$, with associated normal mode
frequencies, $\omega_k$.  A general class of Schr\"{o}dinger-like
equations governing the dynamics of fields in AdS with all their
possible boundary conditions at spatial infinity, were analyzed in
\cite{ishibashiwald}.  Here, we provide the analysis for
the Proca field equation given in Eq.~\eqref{eq:eqdifnormalmodes}
and its solutions.  We
start by
studying the behavior of the solutions near the origin and near
spatial infinity.
We first deal with the behavior at $r=0$.
One can pinpoint the boundary conditions
by checking when the $q_k$ are square-integrable in the sense of $\int
{\bar q}_k q_kdr_*$, where ${\bar q}_k$ means complex conjugate
of $q_k$.
This implies that the regularity conditions are such
that $q_k \propto r^{s}$ as $r\rightarrow 0$,
for some $s$ with $s > -\frac12$. The
functions $q_k$ near $r=0$ have the behavior
\begin{align}\label{eq:r0behaviorPROCA}
q_k =\alpha^{r=0}_k r^{j_k + \frac{d-2}{2}}
+ \beta^{r=0}_k r^{-j_k -\frac{d-4}{4}}\,\,,
\end{align}
where $\alpha^{r=0}_k$ and $\beta^{r=0}_k$ are constants,
and where the equality is valid in
first order near $r=0$.
In all the cases except
the case
$d=4$, $k=2$, and $j_2 = 0$, 
the functions $q_k$ are not square integrable if
$\beta^{r=0}_k$ is finite. 
Therefore,
the regularity condition is $\beta^{r=0}_k=0$. 
In the case of $d=4$, $k=2$, and $j_2=0$,
the function $q_2$ is square integrable if
$\beta^{r=0}_2$ is finite, since $q_2 = \alpha^{r=0}_2
r + \beta^{r=0}_2$.  Still, this solution means that the Proca
field goes as $A_\mu \sim \frac{1}{r}$
and it is rather a solution to the Proca
equations but with a delta dirac distribution as a source term, see
Appendix~\ref{sec:app2}. For
this reason, this particular solution cannot be considered and the
regularity condition $\beta^{r=0}_k= 0$ is maintained.  Such
argument for the regularity condition is also present 
for the scalar field in AdS$_4$ with $\ell = 0$,
see \cite{ishibashiwald}. Note
also that in the case
of the scalar-type Proca field with $d=4$ and $j_2 = 0$, the
asymptotic expansion at $r=0$ of the integrand of the energy 
$E$ defined as
$E =
\int_t T_{\mu\nu} \xi^\mu t^\nu
\frac{r^{d-2}}{\sqrt f} dr d\Omega$, for a
constant $t$ slice,
where $t^\mu$ is its normal vector, $\xi^\mu$
is the timelike Killing vector,
and  $d\Omega$
is the line element of the unit 2-sphere,
seems to have divergent terms
$r^{-1}$, $r^{-2}$ and $r^{-4}$, which only vanish if
$\beta^{r=0}_2 = 0$.

Now, we analyze the boundary conditions at spatial infinity,
$r\to\infty$.  To do that, one looks at the behavior of the Proca
field near $r\rightarrow+\infty$. For the functions $q_k$ to be square
integrable they must behave as $q_k \sim r^s$, for some $s$ with $s
< \frac{1}{2}$.  From the Proca field equations, we get a behavior of
the $q_k$ near $r\rightarrow +\infty$ as
\begin{align}\label{eq:infbehaviorPROCA}
q_k &= \alpha^{r=\infty}_k r^{-
\frac{1}{2}(1 + \sqrt{(d-3)^2 + 4\mu^2l^2})}  
\notag \\&
+ \beta^{r=\infty}_k r^{-\frac{1}{2}(1 - \sqrt{(d-3)^2 +
4\mu^2l^2})} \,\,,
\end{align}
where $\alpha^{r=\infty}_k$ and $\beta^{r=\infty}_k$ are constants,
and where the equality is valid in
first order near $r=\infty$.
For the case $d>4$, the functions $q_k$ are only square integrable if
$\beta^{r=\infty}_k = 0$, which is the Dirichlet boundary condition.
For $d=4$ and $(\mu l)^2 \geq \frac{3}{4}$, the same rationale
applies.  However, for $d=4$ and $0 <(\mu l)^2 < \frac{3}{4}$, the
functions $q_k$ are square integrable for finite $\beta^{r=\infty}_k$.
This is an interesting case because the potential does diverge for
positive $\mu^2$. According to \cite{ishibashiwald}, it is then possible to
impose a one parameter family of boundary conditions. Still, by the
calculation of the usual energy of the Proca field, the condition
$\beta^{r=\infty}_k = 0$ is the only condition that ensures that the
energy is finite and does not diverge on $t$ constant
slices. Although the usual definition of the energy
has been chosen, we
note that there are different valid definitions of the energy function
for the fields
where the energy is finite and
conserved, see \cite{ishibashiwald}.
Nevertheless, we admit Dirichlet boundary conditions for
the $q_k$, which means that
$q_k(r\rightarrow +\infty) = 0$ for every possible
case.

\subsection{Solutions of the Proca equations and normal
mode frequencies}

The Proca equations in Eq.~\eqref{eq:eqdifnormalmodes} can be put in
the form
\begin{align}\label{eq:difrstarnormalmodes}
\partial_{r_*}^2 q_k + \left(\omega_k^2 -
\frac{G_{k}}{\sin^2\left(\frac{r_*} {l}\right)} -
\frac{H_{k}}{\cos^2\left(\frac{r_*}{l}\right)} \right)q_k =
0\,,
\end{align}
where $k\in\{1,2,3\}$,
$r_* = l \arctan\left(\frac{r}{l}\right)$, $G_{k} =
\frac{4j_k(j_k + d -3) + (d-2)(d-4)} {4l^2}$ and $H_{k} =
\frac{(d-2)(d-4) + 4\mu^2l^2}{4l^2}$.  This is a second-order partial
differential equation which is linear and it has three regular
singularity points at  $\frac{r_*}{l} = - \frac{\pi}{2}$,
$\frac{r_*}{l} = 0$, and $\frac{r_*}{l}=+
\frac{\pi}{2}$. Therefore, this
Fuchsian differential equation can be transformed into an
hypergeometric equation, see details in Appendix~\ref{sec:app2}. The
solutions of Eq.~\eqref{eq:difrstarnormalmodes} that satisfy the
regularity conditions at $r=0$ and the Dirichlet boundary conditions
at $r\rightarrow +\infty$ are
\begin{align}\label{eq:solnormalmodes}
&q_k = a_k \left(\frac{r}{l}\right)^{j_k + \frac{d-2}{2}}
\left(1 + \frac{r^2}{l^2}\right)^{n -
\frac{\omega_k l}{2}}\notag\\
&\times\prescript{}{2}{F}_1\left[-n + \omega_k l,-n,j_k
+ \frac{d-1}{2}; 
\frac{\frac{r^2}{l^2}}{1+\frac{r^2}{l^2}}\right]\,\,,
\end{align}
where  $k\in\{1,2,3\}$,
$a_k$ is a constant, and with the normal mode frequencies
$\omega_k $ being
\begin{align}\label{eq:normalmodefreqProca}
\omega_k l = 2 n + j_k + \frac{d-1}{2} 
+ \frac{1}{2}\sqrt{(d-3)^2 + 4\mu^2 l^2}\,\,, 
\end{align}
where $n \in \mathbb{N}_0$, $j_1 = \ell + 1$ with $\ell \in
\mathbb{N}_0$, $j_2 = \ell -1$ with $\ell\in \mathbb{N}$, and $j_3 =
\ell$ with $\ell\in \mathbb{N}$.  Notice that the monopole case of the
Proca field is described by $q_1$ with $j_1 = 1$ or $\ell =
0$. Setting $d=4$, the expression of the normal mode frequencies
given in Eq.~\eqref{eq:normalmodefreqProca}
agrees with the expression given in \cite{tf}.

Although we have not analyzed the case of negative $\mu^2 l^2$, 
the asymptotic behavior in Eq.~\eqref{eq:infbehaviorPROCA} and 
the mode frequencies in Eq.~\eqref{eq:normalmodefreqProca} 
indicate that the eigenvalue problem is well-defined
even for negative field masses, as along as
they obey the inequality,
\begin{align}\label{eq:bfbound}
(\mu l)^2\geq -\frac{(d-3)^2}{4}\,.
\end{align}
The bound given in Eq.~\eqref{eq:bfbound}
is the Proca field analog of the 
Breitenlohner-Freedman bound.
In $d=4$ the
Breitenlohner-Freedman bound for Proca fields in
AdS$_4$ is
$(\mu l)^2 \geq -\frac{1}{4}$.
The Breitenlohner-Freedman bound
was found originally for a massive scalar field in pure AdS
and is given by
$(\mu l)^2 \geq -\frac{9}{4}$
in $d=4$ \cite{Breitenlohner:1982},
and is given by
$(\mu l)^2 \geq -\frac{(d-1)^2}{4}$ for generic $d$
\cite{ishibashiwald}.

\section{Normal modes of the Maxwell field in AdS$_d$
as the $\mu = 0$ limit of
Proca normal modes}
\label{sec:maxwell}

\subsection{Maxwell equations and boundary conditions}

\subsubsection{Maxwell equations}

Maxwellian electromagnetic perturbations, also called Maxwell
perturbations, can be viewed as the limit of Proca perturbations when
$\mu = 0$. However, in Maxwell theory, the field equations are gauge
invariant and the field loses one physical degree of freedom, which
becomes a pure gauge one. Thus, one cannot simply set $\mu = 0$ for
$A_\mu$ in the results above, as one of the degrees of
freedom becomes
spurious. Indeed, the identity on the field $A_\mu$,
Eq.~\eqref{eq:bianchi}, no longer follows directly from the field
equations, becoming simply a gauge choice.  One finds that the gauge
freedom in $A_\mu$ is only scalar, with the vector-type sector of the
Maxwell field being gauge invariant. We display first the final
equations that are of interest here and then we show how to obtain
them from the Proca equations. 

The Maxwell field is described by $d-1$ components.
The time component 
$q_0(t,r) = q_0(r)e^{-i\omega_{12}t}$,
the scalar-type component $q_{12}(t,r) = q_{12}(r) e^{-i\omega_{12}t}$ 
and the $d-3$ vector-type
components $q_3(t,r) = q_3(r) e^{-i\omega_3 t}$, where $\omega_{12}$ 
is a normal mode frequency of 
$q_{12}$ and $\omega_3$ is a normal mode frequency of $q_3$.
The component $q_0$ is given by 
\begin{equation}
q_0 (r) = \frac{i}{\omega_{12} r^{\frac{d}{2}-2}}\partial_{r_*}
\left(q_{12}r^{\frac{d}{2}-2}\right)\,,
\label{q0max}
\end{equation}
and is
completely determined
by the scalar-type component $q_{12}$, which in turn obeys
\begin{equation}\label{eq:maxwell_sch}
\partial_{r_*}^2
q_{12}
+\left(\omega_{12}^2-
V_{12}(r)\right)q_{12} = 0
\,,
\end{equation}
\begin{align}\label{eq:potential_max_scalar}
{V}_{12}(r)=
&f\Bigg[\frac{4\ell(\ell+d-3)+(d-4)(d-2)}{4r^2}\notag
\\& \hskip 6.5em +\frac{(d-4)(d-6)}{4l^2}\Bigg] \,.
\end{align}
This scalar-type component of the Maxwell
field has been denominated as $q_{12}$ 
since it is the
corresponding mode to the Proca $q_1$
and $q_2$ 
modes when the mass of the field is zero $\mu=0$.
The vector-type components covers the
same $d-3$ degrees of freedom as in the massive case and so they are
governed by Eq.~\eqref{eq:difrstarnormalmodes}, for $k=3$ and $\mu=0$,
i.e., 
\begin{align}
&\partial_{r_{*}}^2 q_3 + \left(
\omega_3^2 -
V_3\right) q_3=
0\,\,,\label{eq:qEMv}\\
& V_3= f\Bigg[\frac{4 \ell (\ell+d-3) +
(d-2)(d-4)}{4r^2} \notag \\
& \hskip 10.5em + \frac{(d-2)(d-4)}{4l^2}\Bigg]\,\,,
\label{eq:qEMv2}
\end{align}
where these vector-type components of the Maxwell
field have been denominated as $q_3$ 
since they  are the
corresponding modes to the Proca $q_3$
modes when the mass of the field is zero $\mu=0$.
Surely, there is no possibility of confusion, now we
are dealing with Maxwell modes.

Let us derive the above equations.
In Maxwell theory, the field equations are gauge
invariant
which means that
the Proca
field loses one physical degree of freedom
when $\mu=0$.
Thus, in
the perturbed quantities
there is one mode that becomes nonphysical.
In
order to distinguish between the physical degrees of freedom and the
pure gauge ones, it is useful to work with gauge-invariant variables
rather than with $A_\mu$. Under the gauge transformation,
using the notation
of Eqs.~\eqref{eq:gauge1} and~\eqref{eq:gauge2},
the fields transform as
\begin{align}
&{\tilde A}_a
\to {\tilde A}_a+\partial_a h\,, \nonumber\\
&{\hat A}_i= {\hat A}^{(v)}_i +\hat{\nabla}_i {\hat A}^{(s)} \to
{\hat A}^{(v)}_i+\hat{\nabla}_i \left( {\hat A}^{(s)}+h\right)\,,
\label{gauge}
\end{align}
for some gauge function $h$. One sees that the
gauge freedom in $A_\mu$ is only scalar, with the vector-type sector
of the Maxwell field being gauge invariant.
We start to play with the scalar-type components. 
For the scalar-type components, one has
to go back to 
Eqs.~\eqref{eq:eomScalar1} and \eqref{eq:eomScalar2}. Setting $\mu =
0$ in them  one has
\begin{align}
\tilde{\nabla}_b \tilde{\nabla}^{[a} \psi^{b] }
+ (d-2) \frac{\partial_b r}{r}\tilde{\nabla}^{[a} \psi^{b]} 
+\frac{k_s^2}{2r^2}\left(\psi^a-\partial^a \phi\right) = 0 \ ,
\label{eq:aux_maxwell}
\end{align}
\begin{align}
\tilde{\nabla}_b\left(\psi^b-\partial^b\phi\right) 
+ \left(d-4\right)\frac{\partial_b r}{r}
\left(\psi^b-\partial^b \phi\right)  = 0\,.
\label{m2}
\end{align}
This motivates the definition of the field
$\zeta^a$ given by
\begin{align}
\zeta^a = \psi^a-\partial^a \phi \,,
\label{m3}
\end{align}
which is gauge invariant. Indeed, one
can expand
the gauge function $h$ as
$h(y,\theta) = h (y) Y(\theta)$,
so that, under a gauge transformation, one has
$\psi_a \rightarrow  \psi_a+\partial_a h$ and $\phi 
\rightarrow \phi + h$, where the $\vec{k}_s$ indices were
omitted for convenience. 
In terms of $\zeta^a$ the equations of motion,
Eqs.~\eqref{eq:aux_maxwell}
and \eqref{m2} become
\begin{equation}\label{eq:maxwell_1}
2\tilde{\nabla}_b \tilde{\nabla}^{[a} \zeta^{b]} 
+ 2 (d-2) \frac{\partial_b r}{r}\tilde{\nabla}^{[a} \zeta^{b]}
+\frac{k_s^2}{r^2}\zeta^a= 0 \,,
\end{equation}
\begin{equation}\label{eq:maxwell_2}
\tilde{\nabla}_b\left(r^{d-4} \zeta^b\right)  = 0\,.
\end{equation}
Note that this transformation completely removes a pure gauge degree
of freedom from the system, as $\{\psi^a,\phi\} \rightarrow
\{\zeta^a\}$. This only happens in the massless case, where $\zeta^a$
factorizes. In the background of AdS$_d$ spacetime, substituting
Eq.~\eqref{eq:maxwell_2} in the $r$ component of
Eq.~\eqref{eq:maxwell_1}, and further making the transformation
\begin{align}
q_{12}(r)=\frac{\zeta_r(t,r)}{r}  f
r^{\frac{d}{2}-1}e^{i\omega_{12} t} \,,
\label{t12}
\end{align}
yields the equations,
$\partial_{r_*}^2
q_{12}
+\left(\omega_{12}^2-
V_{12}(r)\right)q_{12} = 0$, 
with
${V}_{12}(r)=
f\Bigg[\frac{4\ell(\ell+d-3)+(d-4)(d-2)}{4r^2}
+\frac{(d-4)(d-6)}{4l^2}\Bigg]$, which
correspond to Eqs.~\eqref{eq:maxwell_sch}
and \eqref{eq:potential_max_scalar}.
Making the transformation
\begin{align}
q_0(r)=
\frac{\zeta_t(t,r)}{r} 
r^{\frac{d}{2}-1}e^{i\omega_{12} t} \,,
\label{t0}
\end{align}
yields
$q_0 (r) = \frac{i}{\omega_{12} r^{\frac{d}{2}-2}}\partial_{r_*}
\left(q_{12}r^{\frac{d}{2}-2}\right)$, 
which
corresponds to Eq.~\eqref{q0max}
and is
completely determined from
the scalar-type component $q_{12}$.
Finally, since the gauge freedom 
in $A_\mu$ is only scalar,
with the vector-type sector of the Maxwell field being 
gauge invariant, see Eq.~\eqref{gauge},
this means that, in
the $\mu = 0$ limit, the vector-type component of $A_\mu$ covers the
same $d-3$ degrees of freedom as in the massive case and so they are
governed by Eq.~\eqref{eq:difrstarnormalmodes}, for $k=3$ and $\mu=0$,
i.e.,
$\partial_{r_{*}}^2 q_3 + \left(
\omega_3^2 -
V_3\right) q_3=
0$
with
$V_3= f\Bigg[\frac{4 \ell (\ell+d-3) +
(d-2)(d-4)}{4r^2}+ \frac{(d-2)(d-4)}{4l^2}\Bigg]$,
which corresponds to Eqs.~\eqref{eq:qEMv}
and \eqref{eq:qEMv2}.

All this agrees with the Maxwell field having $d-2$ degrees of
freedom. Indeed, while the vector-type component covers $d-3$ degrees
of freedom, the scalar-type component only covers one degree of
freedom, which in this case was chosen to be $\zeta_r(t,r)$ or
$q_{12}(r)$. Note also that we are treating only
dynamical solutions of the Maxwell's equations, and so
the frequencies $\omega$ must
be nonzero.

\subsubsection{Boundary conditions}

In summary, the Maxwell field is comprised of
a scalar-type component
$q_{12}$ that satisfies the Maxwell equation
Eq.~\eqref{eq:maxwell_sch}.
and of 
vector-type components
$q_{3}$ that satisfy the Maxwell equation
Eq.~\eqref{eq:qEMv}.

With respect to the regularity conditions, both
$q_{12}$ and $q_{3}$ have the same
behavior as the Proca
$q_k$ in Eq.~\eqref{eq:r0behaviorPROCA} near $r=0$.
Thus,
\begin{align}
q_i=\alpha^{r=0}_i
 r^{\ell + \frac{d-2}{2}}
+ \beta^{r=0}_i
r^{-\ell - \frac{d-4}{2}}\,\,,
\end{align}
where now $i \in \{12,3\}$,
and $\alpha^{r=0}_i$ and $\beta^{r=0}_i$
are constants.  For all cases of
the electromagnetic field, the $q_i$
are square integrable if
 $\beta^{r=0}_i= 0$, and so we admit this condition
as the regularity condition.

In relation to the boundary conditions at spatial infinity,
some care must be taken.
In the massive case, both  scalar-type
and vector-type
perturbations have the same effective mass. This contrasts with the
massless case, where the effective mass for the
the scalar-type component, $\mu_{\mathrm{eff}}^2 =
\frac{(d-4)(d-6)}{4l^2}$, and 
the effective mass for the 
vector-type component,
$\mu_{\mathrm{eff}}^2 = \frac{(d-2)(d-4)}{4l^2}$,
are functionally different, being only the
same in $d=4$. This means that the asymptotic behavior at spatial
infinity is different for the
scalar-type component and for the
vector-type components. Near spatial infinity, the electromagnetic
fields behave as
\begin{align}\label{eq:prd_cases}
q_{12}& = \begin{cases}
\dfrac{\alpha_{12}^{r=\infty}}{r}
+
\beta_{12}^{r=\infty} & \mathrm{for}\,\,d=4\,\,,\\
\\
\dfrac{\alpha_{12}^{r=\infty}}{\sqrt r}
+
\dfrac{\beta_{12}^{r=\infty}\ln r}{\sqrt r} & \mathrm{for}\,\,d=5\,\,,\\
\\
%
\dfrac{\alpha_{12}^{r=\infty}}{r^{\frac{d}{2}-2}}
+
\beta_{12}^{r=\infty}r^{\frac{d}{2}-3}
& \mathrm{for}\,\,d\geq6\,\,,\\
\end{cases}
\\\nonumber
\\
q_3 &= 
\dfrac{\alpha_{3}^{r=\infty}}{r^{\frac{d}{2}-1}}
+
\beta_{3}^{r=\infty}r^{\frac{d}{2}-2} \ \ \ \mathrm{for}\,\,d\geq4
\,\,,
\end{align}
where $\alpha_i^{r=\infty}$ and
$\beta_i^{r=\infty}$ are constants, with $i\in
\{12,3\}$.

For the scalar-type perturbation, $q_{12}$, in $d=4$ the field is
square integrable for nonzero and finite $\beta_i$ when $r\to\infty$.
The requirement that the usual definition of the energy is finite and
independent of the $t$ constant slice allows for the two reflective
boundary conditions, the Dirichlet and the Neumann, with the
Dirichlet imposing $\beta_i^{r=\infty}=0$
and the Neumann imposing
$\alpha_i^{r=\infty}=0$.
Also, other boundary conditions for these
cases are also possible, see \cite{ishibashiwald,Herdeiro2}. Nevertheless, we
impose $\beta_i^{r=\infty}=0$, which corresponds to the Dirichlet boundary
condition.
For the scalar-type perturbation, $q_{12}$, in $d=5$, the field is
square integrable for every $\alpha_{12}^{r=\infty}$,
$\beta_{12}^{r=\infty}$. Moreover, the Dirichlet boundary condition,
which imposes the field to vanish at $r\rightarrow +\infty$, does not
restrict the asymptotic coefficients and leaves the eigenvalue problem
ill-defined.  In order to
have well-defined dynamics for the field in the sense of \cite{ishibashiwald},
one needs to choose 
$\alpha_{12}^{r=\infty}$
and $\beta_{12}^{r=\infty}$ carefully. The
boundary condition that keeps the usual definition of the energy 
to be finite and time independent 
is that $\beta_{12}^{r=\infty} = 0$, which
removes the dominant logarithmic term in Eq.~\eqref{eq:prd_cases}, and
in this case, since
it involves the field and first derivatives
of the field, is a Dirichlet-Neumann boundary condition.
For the scalar-type perturbation, $q_{12}$, in $d=6$, the field is
square integrable for nonzero and finite $\beta_i^{r=\infty}$, as in the case
$d=4$.  Here, it seems that the requirement that the usual definition
of the energy is finite and independent of the $t$ constant slice only
allows the Dirichlet boundary condition.  Also notice that other
boundary conditions for this cases are also possible, see
\cite{ishibashiwald,Herdeiro2}.
Nevertheless, we impose $\beta_i^{r=\infty} = 0$, which
corresponds to the Dirichlet boundary condition.
For the scalar-type perturbations, $q_{12}$, in $d\geq 7$, the field
is square integrable only if $\beta_i^{r=\infty}=0$.
Therefore, Dirichlet
boundary conditions must be imposed.

For the vector-type perturbation, $q_3$, in $d=4$, the field is square
integrable for nonzero and finite $\beta_i^{r=\infty}$.
The requirement that the usual definition of the energy is finite and
independent of the $t$ constant slice allows also
for the two reflective
boundary conditions, the Dirichlet and the Neumann, with the
Dirichlet imposing $\beta_i^{r=\infty}=0$
and the Neumann imposing
$\alpha_i^{r=\infty}=0$.
As well,
other boundary conditions for these cases are
possible, see \cite{ishibashiwald,Herdeiro2}.  Nevertheless, we impose
$\beta_i^{r=\infty} = 0$, which corresponds to the Dirichlet boundary
condition.  For the vector-type perturbations, $q_{3}$, in $d\geq 5$,
the fields are only square integrable if $\beta_i^{r=\infty} = 0$.
Therefore, Dirichlet boundary conditions must be imposed.

With these considerations and for consistency in order to compare with
the Proca field normal modes, we still apply
to the Maxwell field the Dirichlet boundary
conditions for both vector-type and scalar-type perturbations, for
every case, i.e., $q_i(r\to\infty) = 0$.

\subsection{Solutions of the Maxwell equations and normal
mode frequencies}

To obtain the normal modes, one can put the equations
obeyed by $q_{12}$ and  $q_3$ in the form
\begin{align}\label{eq:difeqstarEM}
\partial_{r_*}^2 q_i  +
\left(
\omega_i^2
- \frac{G_i}{\sin^2\left(\frac{r_*}{l}\right)}
- \frac{H_i}{\cos^2\left(\frac{r_*}{l}\right)}
\right)
q_i = 0 \,,
\end{align}
where $i\in \{12,3\}$, and the constants are 
$G_i = \frac{4\ell(\ell+d-3) + (d-2)(d-4)}{4l^2}$, 
$H_{12} = \frac{(d-4)(d-6)}{4l^2}$,
and $H_3 = \frac{(d-2)(d-4)}{4l^2}$.

Imposing the regularity and the
Dirichlet boundary conditions for $d=4$ and $d\geq 6$, and 
Dirichlet-Neumann boundary conditions for $d=5$, one obtains
that the solutions are described by the hypergeometric functions, see
Appendix~\ref{sec:app2}, and they yield,
\begin{align}
q_i = &a_i \left(\frac{r}
{l}\right)^{\ell +\frac{d-2}{2}} 
\left(1+\frac{r^2}{l^2} \right)^{n-
\frac{\omega_il}{2}}\notag \\
&\times \prescript{}{2}{F}_1\left[-n + \omega_il,
-n, \ell + \frac{d-1}{2}; \frac{\frac{r^2}{l^2}}{1 +
\frac{r^2}{l^2}}\right],
\end{align}
for
$i\in \{12,3\}$, with the normal mode frequencies
\begin{align}
\omega_{12} =&
\begin{cases}
2n + \ell + 2 & \mathrm{for}\,\,d=4\,,\\
2n + \ell + 2 & \mathrm{for} 
\,\, d=5 \,\,,\\
2n + \ell + d-3 & \mathrm{for}\,\,d\geq6\,,
\end{cases}
\label{eq:normalfreqEMscalar}\\
\omega_3 = &\quad 2n + \ell + d-2\,\,\,\;\; \mathrm{for}
\,\,d\geq 4\,,\label{eq:normalfreqEMvector}
\end{align}
with $n\in \mathbb{N}_0$ and $\ell \in \mathbb{N}$, where
$\mathbb{R}^+$ are the positive real numbers.

It is interesting to reflect on the normal modes of the
Maxwell
electromagnetic field as the massless limit of the Proca field.
The scalar-type perturbations need to be treated with some care
as we have seen. 
For $d=4$,
the scalar-type perturbation has the same
frequency as the vector-type. Since the massless limit of the Proca
field yields different frequencies for the scalar and vector
perturbations, this discrepancy indicates some discontinuity in the
massless limit, which indeed happens in the frequencies of the scalar
perturbations.
For $d=5$, the normal mode frequencies of the
scalar-type perturbation of the electromagnetic 
field
follow from the massless limit of the 
Proca field. It must be 
noted however that the frequencies for the scalar-type perturbation 
of the electromagnetic field were obtained using Dirichlet-Neumann 
conditions, rather than
Dirichlet boundary conditions that were imposed
to the original  
Proca field. 
Indeed, the Dirichlet boundary conditions leave the eigenvalue problem 
ill-defined.
For $d\geq6$, the electromagnetic scalar-type
normal modes follow
from the massless limit of the modes in
Eq.~\eqref{eq:normalmodefreqProca} for $k=2$ and $j_2 = \ell-1$,
indicating that $q_2$ describes the electromagnetic mode, but now
extended to the massive case, and that $q_1$ describes the scalar
field degree of freedom of the Proca fields, which in the massless
case can be removed by the gauge freedom.
On the other hand,
the
vector-type perturbation of the electromagnetic follows directly from
the massless limit of the vector-type perturbation of the Proca field
and so one sets directly $\mu = 0$ in
Eq.~\eqref{eq:normalmodefreqProca} for $k=3$ and $j_3 = \ell$, and
obtains the normal frequencies in
Eq.~\eqref{eq:normalfreqEMvector}.

We have obtained the normal modes for the Maxwell
electromagnetic perturbations by working in all detail
the massless limit
of the Proca perturbations.
Our results agree
with those found in \cite{ishibashiwald},
where a direct analysis of the Maxwell field from
a master equation in AdS$_d$ was performed. 
The results also conform 
to the Maxwell
electromagnetic perturbations for AdS$_d$
found in \cite{natario,ortega}, and
they recover the 
$\mu = 0$ limit of the
Proca field for AdS$_4$ \cite{tf},
see also \cite{ckjpsl}.

\section{Conclusions}
\label{sec:concl}

In this work, the normal modes of linear Proca perturbations in
AdS$_d$ background were obtained analytically, using the
Ishibashi-Kodama formalism. The Proca field was decomposed into
different components according to its tensorial behavior on the
sphere, yielding scalar-type components, covering two degrees of
freedom of the field, and vector-type components, covering the
remaining $d-3$ degrees of freedom.  In general, while in the
scalar-type sector the two degrees of freedom are coupled, due to the
mass of the field, in the vector-type sector the modes are completely
decoupled. In AdS$_d$, the scalar-type perturbations can be indeed
decoupled in the equations and they are covered independently by two
fields, $q_1$ and $q_2$.  The vector-type perturbations can be covered
by one field only, $q_3$.

The usual regularity boundary conditions at the center and Dirichlet
boundary conditions at infinity were imposed to the Proca equations.
We used a Dirichlet condition because it is the condition that
ensures that the
usual definition of the energy through the conservation of $\xi^\mu
T_{\mu \nu}$, where $\xi^\mu$ is the timelike Killing vector and
$T_{\mu \nu}$ is the stress-energy tensor, is finite and is
independent of the $t$ constant hypersurface.  It would be interesting
to investigate the eigenfrequency problem for other possible boundary
conditions.  Using the regularity and the Dirichlet conditions, the
solutions for the Proca field were obtained and found to be described
by hypergeometric functions. The normal mode frequencies were obtained
for any value of the Proca mass $\mu$ and for $d\geq 4$. For the $d=4$
case, the expression of the frequencies agrees with previous works. 
We  have also found
an explicit expression for Breitenlohner-Freedman bound 
of the Proca field in AdS$_d$.

The normal modes of Maxwell electromagnetic perturbations in 
AdS$_d$ were 
also obtained analytically
through the $\mu = 0$ 
limit of Proca perturbations. In this case, by working with gauge-invariant
variables, it was possible to separate the physical modes from the
nonphysical ones. For consistency, the regularity and Dirichlet 
boundary conditions were used, except for the case of $d=5$ where 
the Dirichlet-Neumann boundary condition was used, in order to 
analyze the massless limit of the Proca field. We
have thus
recovered the normal mode frequencies
of the Maxwell field found in previous works,
with the $d=4$ case regarding the scalar perturbations 
having to be treated
with much care, in particular, the massless limit of the 
Proca scalar perturbations in $d=4$ does not lead directly to the 
Maxwell field scalar perturbation modes.

\acknowledgments{
We thank Antonino Flachi for conversations.  We acknowledge financial
support from Funda\c{c}\~ao para a Ci\^encia e Tecnologia - FCT
through the project No. UIDB/00099/2020 and project
No. UIDP/00099/2020. TF acknowledges a grant from FCT
no.~RD0970.
}

\appendix
\section{Spherical harmonics on the ($d-2$)-sphere}
\label{sec:app1}

\subsection{Initial considerations}

The decomposition of fields in spherical harmonics is
important to study
the structure of their perturbations and to isolate
the physical degrees
of freedom of the fields themselves.
Here, we discuss in some detail the
properties of these special functions.

The approach we adopt here
to construct 
spherical harmonics on the $(d-2)$-sphere, $\mathcal{S}^{d-2}$, 
follows \cite{myers}, 
where spherical harmonics are constructed by embedding $\mathcal{S}^{d-2}$ 
in a ($d-1$)-Euclidean space, $\mathbb{R}^{d-1}$. 
Firstly, we introduce some useful concepts: (a) a polynomial $h_\ell:
\mathbb{R}^{d-1} \rightarrow \mathbb{C}$ is homogeneous of degree
$\ell$ in $\mathbb{R}^{d-1}$ if $h_\ell(\lambda x^\mu) = \lambda^\ell
h_\ell (x^\mu)$, for any $\lambda \in \mathbb{R}, {x^\mu} \in
\mathbb{R}^{d-1}$; (b) a polynomial $h_\ell: \mathbb{R}^{d-1}
\rightarrow \mathbb{C}$ is harmonic in $\mathbb{R}^{d-1}$ if $\Box_{E}
h_\ell = 0$, where $\Box_{E}$ is the Laplacian on $\mathbb{R}^{d-1}$;
(c) a spherical harmonic of degree $\ell$ on $\mathcal{S}^{d-2}$ is a
function $Y_\ell: \mathcal{S}^{d-2} \rightarrow \mathbb{C}$ such that,
for some homogeneous and harmonic polynomial in $\mathbb{R}^{d-1}$,
$h_\ell$, $Y_\ell(\theta)=h_\ell(\theta)$ for all 
$\theta = (\theta^2,...\theta^{d-1}) \in
\mathcal{S}^{d-2}$.
Spherical harmonics on the $(d-2)$-sphere can also
be constructed 
recursively by dimensional reduction, see, e.g.,
\cite{higuchi}, but
the approach becomes cumbersome when constructing 
vector spherical harmonics.
For more details on the matter we refer 
to \cite{lovenumbers,myers,higuchi,abramowitz}, see
also \cite{ishibashi_kodama2}.

\subsection{Scalar spherical harmonics}

In spherical coordinates, the $\mathbb{R}^{d-1}$ line element, $ds^2$,
is related to the $\mathcal{S}^{d-2}$ line element, $(d\Omega^{d-2})^2 =
{\hat g}_{i j}d\theta^id\theta^j$, by $ds^2 = dr^2 +
r^2(d\Omega^{d-2})^2$.  Using Eq.~\eqref{eq:chris}, the nonvanishing
Christoffel symbols associated to $g_{\mu \nu}$ in these coordinates
are
\begin{equation}
\Gamma^r_{\ i j} = -r{\hat g}_{i j} \,,\quad 
\Gamma^i_{\ r j} = \frac{1}{r}\delta^i_j \,, 
\quad \Gamma^i_{\ j k} = \hat{\Gamma}^i_{\ j k},
\end{equation}
and the condition for a homogeneous polynomial of degree $\ell$,
$h_\ell$, to be harmonic in $\mathbb{R}^{d-1}$, becomes
\begin{equation}\label{eq:app_sh}
\Box_{E} h_\ell = \frac{1}{r^{d-2}}\partial_r\left(r^{d-2} 
\partial_r h_\ell\right) + \frac{1}{r^2}\hat{\Box} h_\ell = 0 \,,
\end{equation}
where $\hat{\Box} = {\hat g}^{i j} \hat{\nabla}_i \hat{\nabla}_j$. 
Since $h_\ell$ is homogeneous, it follows that
$h_\ell (x) |_{S} = r^\ell h_\ell ({\hat{x}}) |_{S} = r^\ell 
Y _\ell (\theta)$, where $x^\mu = r\hat{x}^\mu$,
so that, substituting in Eq.~\eqref{eq:app_sh}, one has
\begin{equation}\label{eq:eigenmodes_scalar}
\hat{\Box} Y_{\ell} = -\ell(\ell+d-3) Y_{\ell} \,\, , 
\,\, \ell = 0,1,2,...
\end{equation}
$Y_{\ell}$ are called scalar spherical harmonics. 
Besides being eigenfunctions of $\hat{\Box}$, 
it can also be shown that they form a complete and 
orthogonal set on $\mathcal{S}^{d-2}$.

\subsection{Vector spherical harmonics}

Vector spherical harmonics can be constructed in the same way as
scalar spherical harmonics, only this time one starts with vector
functions $V_{\nu}^\ell: \mathbb{R}^{d-1} \rightarrow
\mathbb{C}^{d-1}$. Using the Helmholtz-Hodge theorem, $V_{\nu}^\ell$
can be written as, in analogy to Eq.~\eqref{eq:gauge1},
\begin{equation}
V_{\nu}^\ell dx^\nu = V_{r}^\ell dr +  (W_i^\ell
+\hat{\nabla}_i \sigma^\ell) d\theta^i \,\,, 
\,\, \hat{\nabla}^i W_i^\ell = 0 \,\,,
\end{equation}
where $W^i_\ell$ is a vector on the $(d-2)$-sphere and 
$V_r^\ell$ and $\sigma^\ell$ are scalars. 
Expanding $\Box_{E} V_\nu^\ell$ in spherical coordinates, 
and assuming that $V_{\nu}^\ell$ is harmonic, one has
\begin{equation}\label{eq:W_vh}
\Box_{E} W_i^\ell = \partial^2_r W_i^\ell 
+ \frac{d-4}{r} \partial_r W_i^\ell 
- \frac{d-3}{r^2}W_i^\ell 
+ \frac{1}{r^2} \hat{\Box} W_i^\ell = 0 \,,
\end{equation}
as well as two coupled equations for the 
scalars $V_r^\ell$ and $\sigma^\ell$ \cite{myers}. 
Here, we are only interested in the equations for $W_i^\ell$. 
Since $V_\nu^\ell$ is homogeneous, $V_\mu^\ell(x) =
r^\ell V_\mu^\ell(\theta)$. 
This means that
${V}_\nu^\ell(x) dx^\nu =   
\frac{\partial x^\mu}{\partial r} r^\ell V_\mu^\ell(\theta) dr 
+ \frac{\partial x^\mu}{\partial \theta^i} r^\ell
V_\mu^\ell(\theta) d\theta^i $
and so $W^\ell_i + \hat{\nabla}_i \sigma^\ell = 
r^{\ell+1}\frac{\partial \hat{x}^\mu}{\partial \theta^i}
V_\mu^\ell(\theta)$, where 
$x^\mu = r \hat{x}^\mu$.
Hence, we define $Y_{\ell i}(\theta)$ as the angular dependence of 
$W^\ell_i(x)$, i.e.,
$W^\ell_i(x) |_S  = r^{\ell + 1} Y_{\ell\, i}(\theta)$, 
which put in Eq.~\eqref{eq:W_vh}, it follows that
\begin{equation}\label{eq:eigenmodes_t}
\hat{\Box} Y_{\ell\, i} = -\left[\ell(\ell+d-3)-
1 \right] Y_{\ell\, i} \,, \,\, \,\, \ell = 1,2,...
\end{equation}
The vectors $Y_{\ell\, i}$ are called transverse vector spherical
harmonics, as they verify $ \hat{\nabla}^i Y_{\ell\, i} = 0$.  One can
also construct longitudinal vector spherical harmonics, $Y_{\ell\,
i}^L$, by taking the gradient of scalar fields on $\mathcal{S}^{d-2}$.
These are defined as \cite{lovenumbers}
\begin{equation}
Y_{\ell\, i}^L \equiv \frac{1}{\sqrt{\ell(\ell+d-3)}} \hat{\nabla}_{i} 
Y_{\ell} \,,
\end{equation}
and have eigenmodes
\begin{equation}\label{eq:eigenmodes_l}
\hat{\Box} {\hat{\nabla}_i} Y_{\ell} = -\left[\ell(\ell+d-3)-(d-3)\right]
{\hat{\nabla}_i} Y_{\ell} \,\,, \,\, \ell = 1,2,...
\end{equation}

\subsection{Properties under rotation and parity transformations}

Now, let $\varphi$ and $u^i$ be, respectively, a scalar field and a
vector field on $\mathcal{S}^{d-2}$.  The action of the SO($d-1$)
Casimir operator, $\hat{J}^2$, on these two entities is $\hat{J}^2
\varphi = -\Box_S \varphi$ and $\hat{J}^2 u^i = -\left(\Box_S-(d-3)\right)
u^i$, see \cite{ishibashiwald}.  Using
Eqs.~\eqref{eq:eigenmodes_scalar},~\eqref{eq:eigenmodes_t},
and~\eqref{eq:eigenmodes_l}, one gets the following
\begin{align}
&\hat{J}^2 Y_{\ell} = \ell(\ell+d-3) Y_{\ell} \,\, ,\\
&\hat{J}^2 \hat{\nabla}_i Y_{\ell} = 
\ell(\ell+d-3) \hat{\nabla}_i Y_{\ell} \,\, ,\\
&\hat{J}^2 Y_{\ell\, i} = \left[\ell(\ell+d-3)+d-4\right] 
Y_{\ell\, i} \,\, .
\end{align}
One sees that, for $\ell \geq 1$ and $d > 4$, the Casimir eigenvalues
of the scalar spherical harmonics and of the longitudinal vector
spherical harmonics are the same.  One then expects these modes to
mix. On the contrary, the Casimir eigenvalues of the transverse vector
spherical harmonics are never equal for $d>4$ to the eigenvalues of
the longitudinal vector spherical harmonics, so that these completely
decouple.
For $d=4$, the last are equal and one might expect to have
mixed modes.  However, in this case, the modes are decoupled due to
their different parity eigenvalues. Indeed, under parity
transformations
$\theta_{i = 2} \rightarrow \pi-\theta_{i = 2}$ and 
$\theta_{i =3} \rightarrow \pi+\theta_{i= 3}$, one has 
$\hat{\mathcal{P}}Y_{\ell} = \left(-1\right)^\ell Y_{\ell}$, 
$\hat{\mathcal{P}} \hat{\nabla}_i Y_{\ell} = 
\left(-1\right)^\ell \hat{\nabla}_i Y_{\ell}$
and $\hat{\mathcal{P}}Y_{\ell\, i} = 
\left(-1\right)^{\ell+1} Y_{\ell\, i}$.
Note that a vector, $A_{\theta_{i}}$, on the $(d-2)$-sphere transforms
under parity as $A_{\theta_{i=1}} \rightarrow A_{\theta_{i=1}}$ and
$A_{\theta_{i \neq 1}} \rightarrow -A_{\theta_{i \neq 1}}$.

\section{Hypergeometric differential equation}
\label{sec:app2}

\subsection{General equation}

We study now the solutions of
the hypergeometric differential 
equation, based on  \cite{abramowitz}, see also \cite{ishibashiwald}.

The Proca and electromagnetic field equations can be separated and
decoupled to assume the reduced form
\begin{equation}\label{eq:schrodingerapp}
\partial_{r_*}^2 q + \left(\omega^2 -
\frac{G}{\sin^2\left(\frac{r_*}{l}\right)} -
\frac{H}{\cos^2\left(\frac{r_*}{l}\right)} \right)q = 0 \,,
\end{equation}
as it appears in Eqs.~\eqref{eq:difrstarnormalmodes}
and~\eqref{eq:difeqstarEM}, where $r_* = l
\arctan\left(\frac{r}{l}\right)$,
$
G =\frac{4j (j +d-3)+(d-2)(d-4)}{4l^2}$
for $j\in \mathbb{N}_0$,
and $H$ for now is some constant
that depends on $d$, $l$, $\mu$, and $j$
with units of $\frac1{l^2}$.
By making the change
\begin{equation}
q(z) = z^\alpha (1-z)^\beta \Theta(z) \,,
\end{equation}
with  
\begin{equation}
z = \sin^2\left(\frac{r_*}{l}\right)\,,
\end{equation}
and recalling $r_*=l\arctan\frac{r}{l}$,
one can
transform Eq.~\eqref{eq:schrodingerapp}
into the hypergeometric differential equation
\begin{equation}\label{eq:hypergeometric}
z(1-z) \frac{d^2\Theta}{dz^2} + \left[c -(a +b +1)z\right]
\frac{d\Theta}{dz} - a b \Theta = 0 \,,
\end{equation}
with $a = \alpha + \beta + \frac{\omega l}{2}$, $b = \alpha + \beta -
\frac{\omega l}{2}$, $c = 2\alpha + \frac{1}{2}$, and
$
\alpha = \frac{2j+d-2}{4}
\label{eq:appalphaexpression}
$,
$\beta = \frac{1}{4}\left(1+\sqrt{1 + 4H l^2}\right)
\label{eq:appbetaexpression}
$.
For $d$ even,
the solutions of (\ref{eq:hypergeometric}) 
are given in terms of the hypergeometric function, 
$\prescript{}{2}{F}_1$, as  \cite{abramowitz}
\begin{align}\label{eq:solution} &q(z) = A z^{\alpha} (1-z)^\beta
\prescript{}{2}{F}_1\left[a,b,c;z\right] \notag \\
&+ B z^{1/2-\alpha} (1-z)^\beta \prescript{}{2}{F}_1\left[a-c+1,b -c
+1,2-c ;z\right] \notag\\
&\hskip 15.8em \mathrm{for}\,\, d\,\,\mathrm{even}\,,
\end{align}
where $A$ and $B$ are constants of integration. For $d$ odd,
the indicial roots of Eq.~\eqref{eq:hypergeometric} are separated by
an integer, making the solutions presented linearly dependent. In this
case, since $c\geq 2$ and it is an integer, $q(z)$ can be written as
\cite{abramowitz}
\begin{align}\label{eq:solutionodd}
&q(z) = A z^{\alpha} (1-z)^\beta
\prescript{}{2}{F}_1\left[a,b,c;z\right]
+ B z^{\alpha } (1-z)^\beta \notag \\
&\times\Bigg[\prescript{}{2}{F}_1
\left[a,b,c;z\right]\ln{z} + 
\sum^\infty_{i=1}v_{i}z^i- \sum^{c-1}_{i=1}w_{i}z^{-i}\Bigg] \notag\\
& \hskip 15.5em \mathrm{for}\,\, d\,\,\mathrm{odd}\,,
\end{align}
where 
\begin{align}
&v_{i} = \frac{(a)_i (b)_i}{(c)_i (i!)}[
\Psi(a+i) -
\Psi(a) + \Psi(b+i) - \Psi(b)\notag\\
&-\Psi(c+i) + \Psi(c) - \Psi(1+i) + \Psi(1)]\,\,,\\
&w_{i} = \frac{(i-1)! (1-c)_i}{(1 - a)_i (1 - b)_i}\,\,,
\end{align}
$(a)_i = \frac{\Gamma(a + i)}{\Gamma(a)}$ if $a>0$,
$(a)_i =
(-1)^i (-a-i+1)_i$ if $a<0$, and
the same for $(b)_i$ and $(c)_i$, and 
$\Psi(a) =
\frac{\Gamma'(a)}{\Gamma(a)}$ is the digamma function, see
\cite{abramowitz}.

\subsection{Imposing regularity conditions near $r=0$, i.e.,
$z=0$}

To obtain the normal modes, one needs to impose boundary conditions
that characterize the system. At $r=0$ or $z=0$, one imposes
regularity conditions so that the solution does not diverge there.

For
even spacetimes, it is seen from Eq.~\eqref{eq:solution}
that for $\alpha>\frac12$ one needs to set $B=0$, considering that
$\lim_{z\to0} \prescript{}{2}{F}_1\left[a,b,c;z\right] = 1$.
For 
$\alpha=\frac12$, which 
corresponds to the case 
$q_2$ with
$\ell=1$ and $d=4$, the exponent in $z$ vanishes and the solution
seems to be finite near $z=0$.  However, this is just an artifact of
having removed the origin when separating the field in spherical
harmonics, see \cite{ishibashiwald}. Indeed, if one writes the components of
the four-dimensional Proca field $A_r$, $A_{\theta_2}$ and
$A_{\theta_3}$ in terms of $q_{2}$ with $\ell = 1$ and the other
fields put to zero, one arrives at
$A_r =\frac{1}{r f}\sum_{m=-1}^{1} q_{2\,(0,m)} Y_{(1,m)}$,
$A_{\theta_2} = \sum_{m=-1}^{1}
q_{2\,(0,m)}\partial_{\theta_2} Y_{(1,m)}$, and
$A_{\theta_3} = -\sum_{m=-1}^{1}$
$\frac{q_{2\,(0,m)}}{d-3}\partial_{\theta_3} Y_{(1,m)}$,
where $q_{2\,(0,m)}$ is $q_2$ with $j_2 = 0$, $\ell=1$ and
azimuthal number $m$. 
Since near the origin, the spacetime is flat, i.e.,
$f(r)\simeq 1$, and $q_{2\,(0,m)} = B_{(1, m)}$ where $B_{(1, 1)}$, 
$B_{(1, 0)}$ and $B_{(1, -1)}$ are constants, the behavior of the
Proca field near the origin 
becomes
$A_r \simeq \frac{1}{r}\left(B_{(1,1)}Y_{(1,1)}+B_{(1,0)}Y_{(1,0)}
+B_{(1,-1)}Y_{(1,-1)}\right)$,
$A_{\theta_2} \simeq B_{(1,1)}\partial_{\theta_2}
Y_{(1,1)}+B_{(1,0)}\partial_{\theta_2} Y_{(1,0)}
+B_{(1,-1)}\partial_{\theta_2} Y_{(1,-1)}$,
$A_{\theta_3} \simeq B_{(1,1)}\partial_{\theta_3}Y_{(1,1)}+
B_{(1,0)}\partial_{\theta_3}Y_{(1,0)}
+B_{(1,-1)}\partial_{\theta_3}Y_{(1,-1)}$.
The components of a vector field transform as 
$A_{\mu'} = \frac{\partial x^\mu}{\partial x^{\mu'}}A_{\mu}$ 
to give $A_x,A_y,A_z \sim \frac{1}{\sqrt{x^2+y^2+z^2}}$, in
Cartesian coordinates. 
Since the Proca field equations Eq.~\eqref{eq:proca_equation} can be
written 
near the origin in Cartesian coordinates as 
$
\partial^\nu\partial_\nu A_\rho -\mu^2 A_\rho = 0$,
where $\alpha=\{t,x,y,z\}$, one has 
$\left(\partial^2_x+\partial^2_y+\partial^2_z\right) A_{x,y,z} 
\sim \delta(x)\delta(y)\delta(z)$.
Due to this additional delta term, $q_{2\,(0,m)} = B_{(1, m)}$, with
$B_{(1, m)} \neq 0$, cannot be a solution near the origin and one
needs to set $B = 0$ in Eq.~\eqref{eq:solution}, even in this special
case. Moreover, if one takes the asymptotic expansion of the integrand
of the usual energy, $T_{tt}\frac{r^{d-2}}{f}$, near $r=0$, then one
gets divergent terms $r^{-1}$, $r^{-2}$ and $r^{-4}$ which only vanish
if $B=0$. Thus, in this particular case of
$\alpha=\frac12$, which 
corresponds to the case 
$q_2$ with
$\ell=1$ and $d=4$, one has also $B=0$.

For odd spacetimes, all the terms of Eq.~\eqref{eq:solutionodd} 
except the last one vanish in the limit $z \rightarrow 0$. 
For the last term, as $c\geq2$ and it is an integer, 
it can be seen that $\sum^{c-1}_{i=1}w_{i}z^{\alpha - i}$ 
contains a power $r^s$ with $s\leq - \frac12$ always and so the 
field $q$ would not be square integrable. 
Thus, one also needs to set $B = 0$ in this case.

\subsection{Imposing Dirichlet boundary conditions at spatial
infinity, $r\to\infty$, i.e., $z\to1$}

\subsubsection{Expansion at spatial infinity for general case}

To impose the remaining boundary condition, one uses 
the transformation law $z\rightarrow 1-z$ of $\prescript{}{2}{F}_1$, 
so that, if $c-a-b=\frac12-2\beta \neq -m'$  with
$m' \in \mathbb{N}_0$ one has
\cite{abramowitz}
\begin{align}\label{eq:property_hyper1infexpand}
&q(z) =A z^\alpha (1-z)^\beta \notag\\
&\times\Bigg[
\frac{\Gamma(c)\Gamma(1-e)}{\Gamma(\Bar{a})\Gamma(\Bar{b})} 
\prescript{}{2}{F}_1\left[a,b,e;1-z\right]
\notag\\
&+ (1-z)^{\frac12-2\beta}
\frac{\Gamma(c)\Gamma(e-1)}{\Gamma(a)\Gamma(b)}
\prescript{}{2}{F}_1\left[\Bar{a},\Bar{b},2-e;1-z\right]\Bigg]\,,
\end{align}
where $e = 1 - c + a +b$, $\Bar{a} = c-a$ and $\Bar{b} = c-b$. 
If $c-a-b=\frac12-2\beta = 0$, one has
\begin{align}\label{eq:property_hyper2infexpand}
&q(z) =A z^\alpha (1-z)^\beta \frac{\Gamma(a+b)}{\Gamma(a)\Gamma(b)}
\sum_{i=0}^\infty \frac{(a)_i (b)_i}{(i!)^2}(1-z)^i\notag\\
& \times \left[2\Psi(i+1) - \Psi(a+i) - \Psi(b+i)
- \log(1-z) \right]\,.
\end{align}
If $c-a-b=\frac12-2\beta = -m'$ with $m' \in \mathbb{N}$, one has
\begin{align}\label{eq:property_hyper3infexpand}
&q(z) = A z^\alpha (1-z)^\beta \Bigg(\frac{\Gamma(m')
\Gamma(c)}{\Gamma(a)\Gamma(b)} 
\sum^{m'-1}_{i=0} v'_i (1-z)^{i-m'}\notag \\
&-(-1)^{m'}\sum^\infty_{i=0}
\frac{ \Gamma(c) (1-z)^i}{\Gamma(a-m')\Gamma(b-m')}
\left[w'_i\ln{(1-z)}+t_i \right]\Bigg) \,\,,
\end{align}
with the coefficients defined by
\begin{align}
& v'_i = \frac{(a-m')_i (b-m')_i}{i! (1 - m')_i}\,\,,\\
& w'_i = \frac{(a)_i (b)_i}{i! (i+m')!}\,\,,\\
& t_i = w'_i [ \Psi(a+i) + \Psi(b+i) \notag \\
&- \Psi(i+1) - \Psi(i+m'+1)]\,\,.
\end{align}

\subsubsection{Proca field case}

In the Proca field case, we have three fields $q_1$, $q_2$ and $q_3$,
where the value of $j$ that appears in
Eq.~\eqref{eq:appalphaexpression} for each field is $j_1 = \ell +1$,
$j_2=\ell -1$ and $j_3 = \ell$. The constant $H$ in this case is
$H = \frac{(d-2)(d-4) + 4\mu^2 l^2}{4l^2}$, which leads to the
expression of $\beta$ in Eq.~\eqref{eq:appbetaexpression} to become
$\beta= \frac{1}{4}(1 + \sqrt{(d-3)^2 + 4\mu^2l^2})$. Moreover, we
have $c-a-b \neq -m'$ with $m'\in \mathbb{N}_0$. Therefore the correct
expansion at spatial infinity of the Proca field is described by
Eq.~\eqref{eq:property_hyper1infexpand}. Since $\beta > \frac12$, the
first term of Eq.~\eqref{eq:property_hyper1infexpand} vanishes as the
gamma function in the numerator is finite.  The remaining term must be
zero to satisfy the Dirichlet boundary condition, which only occurs if
either $a$ or $b$ are nonpositive integers. By requiring that $\omega
>0$, the Dirichlet boundary condition leads to $b = -n$, with $n \in
\mathbb{N}_0$, and so the normal mode eigenfrequencies $\omega_k l$
for each field $q_k$ are
\begin{align}
\label{eq:frequencies_app}
&\omega_{k} L = 2n + j_k + \frac{d-1}{2}+
\frac{1}{2}\sqrt{\left(d-3\right)^2+4\mu^2L^2} \,\, , 
\end{align}
where the normal mode eigenfunctions are given by 
\begin{align}
 &q_k (r) = A_k \left(\frac{r}{l}\right)^{\frac{2j_k+d-2}{2}}
 \left(1+\frac{r^2}{l^2}\right)^{n-\frac{\omega_k l}{2}} 
 \notag \\ &\times
 \prescript{}{2}{F}_1\left[-n+\omega_k l,-n,
 j_k+\frac{d-1}{2};\frac{r^2/l^2}{1+r^2/l^2}\right] \,\,.
\end{align}

\subsubsection{Maxwell electromagnetic field case}

\noindent
{\it Scalar-type perturbation:}
\vskip 0.1cm

\noindent
The scalar-type perturbations for the Maxwell
field are described by the function
$q_{12}$, where the value of $j$ that appears in
Eq.~\eqref{eq:appalphaexpression} is $j=\ell$. The constant $H$ in
this case is $H = \frac{(d-4)(d-6)}{4l^2}$, which leads to the
expression of $\beta$ in Eq.~\eqref{eq:appbetaexpression} to become
$\beta= \frac{1}{4}(1 + \abs{d-5})$.

For even dimensions, the expansion at spatial infinity of
$q_{12}$ is described by
Eq.~\eqref{eq:property_hyper1infexpand}, since $c-a-b \neq -m'$, with
$m' \in \mathbb{N}_0$. We distinguish two cases here, $d=4$ and
$d\geq 6$. At spatial infinity for $d=4$, since $\beta=\frac12$,
the first
term in Eq.~\eqref{eq:property_hyper1infexpand} vanishes while the
second term is finite.  We still impose the Dirichlet boundary
conditions and the first term vanishes if $b = -n$, with $n\in
\mathbb{N}_0$ and $\omega l > 0$. Therefore, the normal mode
frequencies are
\begin{align}\label{eq:appnormalmodesEMscalarfreqd4}
\omega_{12} = 2n + \ell + 2\,,\quad\quad\mathrm{for}\,\,d=4\,.
\end{align}
For $d\geq 6$ and even, at spatial infinity, since $\beta \geq
\frac{1}{2}$, the first term in
Eq.~\eqref{eq:property_hyper1infexpand} vanishes while the second term
seems to diverge for $d>6$ and seems to assume a finite value for
$d=6$. Despite the $d=6$ case, we impose the Dirichlet boundary
conditions and the second term vanishes if $b = -n$, with $n\in
\mathbb{N}_0$ and $\omega l > 0$. Therefore, the normal mode
frequencies are
\begin{align}\label{eq:appnormalmodesEMscalarfreqd6}
\omega_{12}  = 2n + \ell + d-3\,,
\quad\quad\mathrm{for}\,\,d\geq6\,.
\end{align}

For odd dimensions, we split the analysis for $d=5$ and $d \geq7$.
For $d=5$, the expansion at spatial infinity of
$q_{12}$ is described by
Eq.~\eqref{eq:property_hyper2infexpand}, since $c-a-b = 0$. Therefore,
at spatial infinity, all the terms vanish in
Eq.~\eqref{eq:property_hyper2infexpand}. The Dirichlet boundary 
conditions, which impose the field to vanish at $r\rightarrow +\infty$, 
makes the eigenvalue problem ill-defined. 
One can instead impose a one parameter family boundary condition. For 
example, 
one can choose a boundary condition such that the logarithmic term 
vanishes
called Dirichlet-Neumann, which
is satisfied if $b=-n$, with $n\in \mathbb{N}_0$, 
and so the frequency is given by
\begin{align}
    \omega_{12}  = 2n + \ell + 2\,,
\quad\quad\mathrm{for}\,\,d = 5\,.
\end{align}
For $d\geq7$, the expansion at
spatial infinity of $q_{12}$ is described by
Eq.~\eqref{eq:property_hyper3infexpand}, with $\beta > \frac{1}{2}$
and $c-a-b = -m'$ with $m'\in \mathbb{N}$. In this case, the second
term vanishes while the first term seems to diverge. We impose the
Dirichlet boundary conditions and so the first term vanishes if $b=-n$
with $n\in \mathbb{N}_0$. The normal mode eigenfrequencies are then
also described by Eq.~\eqref{eq:appnormalmodesEMscalarfreqd6}.

The eigenfunctions for all the even and odd dimension cases are
\begin{align}
&q_{12} (r) = A_{12}
\left(\frac{r}{l}\right)^{\frac{2\ell+d-2}{2}}
\left(1+\frac{r^2}{l^2}\right)^{n-
\frac{\omega_{12} l}{2}} 
\notag \\ &\times
\prescript{}{2}{F}_1\left[-n+\omega_{12} l,
-n, \ell +\frac{d-1}{2};
\frac{\frac{r^2}{l^2}}{1+\frac{r^2}{l^2}}\right] \,.
\end{align}

\noindent
{\it Vector-type perturbation:}
\vskip 0.1cm

\noindent
The vector-type perturbations are described by the function
$q_3$, where the value of $j$ that appears in
Eq.~\eqref{eq:appalphaexpression} is $j=\ell$. The constant $H$ in
this case is $H = \frac{(d-2)(d-4)}{4l^2}$, which leads to the
expression of $\beta$ in Eq.~\eqref{eq:appbetaexpression} to become
$\beta= \frac{1}{4}(d-2)$.

For even dimensions, the expansion at spatial infinity of
$q_3$ is described by
Eq.~\eqref{eq:property_hyper1infexpand}, since $\beta \geq \frac12$ and
$c-a-b \neq -m'$, with $m' \in \mathbb{N}_0$.  At spatial infinity, it
turns out that the first term in
Eq.~\eqref{eq:property_hyper1infexpand} vanishes while the second term
seems to diverge for $d>4$ and assumes a finite value for
$d=4$. Despite the $d=4$ case, we impose Dirichlet boundary conditions
and the second term only vanishes if $b=-n$, with $n\in
\mathbb{N}_0$. Therefore, the normal mode eigenfrequencies are given
by
\begin{align}\label{eq:appnormalmodesEMvectorfreq}
\omega_3 l = 2n + \ell + d-2\,\,. 
\end{align}

For odd dimensions, the expansion at spatial infinity of
$q_3$ is described by
Eq.~\eqref{eq:property_hyper3infexpand}, since $\beta>\frac{1}{2}$ and
$c-a-b = -m'$, with $m' \in \mathbb{N}$. At spatial infinity, the
first term in Eq.~\eqref{eq:property_hyper3infexpand} seems to
diverge. Imposing the Dirichlet boundary conditions, these terms
vanish if again $b=-n$ with $n\in \mathbb{N}_0$. The normal mode
eigenfrequencies for odd dimensions have then the same expression as
Eq.~\eqref{eq:appnormalmodesEMvectorfreq}.

The eigenfunctions for all the  even and odd dimension
cases are
\begin{align}
&q_3 = A_3
\left(\frac{r}{l}\right)^{\frac{2\ell+d-2}{2}}
\left(1+\frac{r^2}{l^2}\right)^{n-
\frac{\omega_3 l}{2}} 
\notag \\ &\times
\prescript{}{2}{F}_1\left[-n+\omega_3
l,-n, \ell +\frac{d-1}{2};
\frac{\frac{r^2}{l^2}}{1+\frac{r^2}{l^2}}\right] \,\,.
\end{align}


\begin{thebibliography}{999}


\bibitem{Calabi:1962}
E.~Calabi and L.~Markus, ``Relativistic space forms",
Ann. Math. {\bf 75}, 63  (1962).



\bibitem{Penrose:1968}
R.~Penrose, ``The structure of space-time'', in 
{\it Battelle
Rencontres 1967 Lectures in Mathematical Physics}, edited by 
B. DeWitt and J. A. Wheeler (Benjamin, New York, 1968),
p. 121.

\bibitem{hawking}
S.~W.~Hawking and G.~F.~R.~Ellis,
{\it The Large Scale Structure of Space-Time},
(Cambridge University Press, Cambridge, 2023).



\bibitem{avis}
S.~J.~Avis, C.~J.~Isham, and D.~Storey,
``Quantum field theory in anti-De Sitter space-time",
Phys. Rev. D \textbf{18}, 3565 (1978).


%
\bibitem{Breitenlohner:1982}
P.~Breitenlohner and D.~Z.~Freedman, 
``Positive energy in anti-de Sitter backgrounds and gauged extended 
supergravity", Phys. Lett. B {\bf 115}, 197 (1982).


\bibitem{maldacena}
J.~M.~Maldacena,
``The large N limit of superconformal field theories and supergravity",
Adv. Theor. Math. Phys. \textbf{2}, 231 (1998);
arXiv:hep-th/9711200 [hep-th].



\bibitem{bizon1}
P.~Bizo\'n and A.~Rostworowski,
``Weakly turbulent instability of anti-de Sitter space",
Phys. Rev. Lett. \textbf{107}, 031102 (2011);
arXiv:1104.3702 [gr-qc].




\bibitem{Buchel:2012}
A.~Buchel, L.~Lehner, and S.~L.~Liebling, 
``Scalar collapse in AdS spacetimes", 
Phys. Rev. D {\bf 86}, 123011 (2012);
arXiv:1210.0890 [gr-qc].


\bibitem{Masachs:2019}
R.~Masachs and B.~Way, ``New islands of stability with
double-trace deformations", 
Phys. Rev. D {\bf 100}, 106017 (2019);
arXiv:1908.02296 [hep-th].


\bibitem{regge_wheeler}
T.~Regge and J.~A.~Wheeler,
``Stability of a Schwarzschild singularity",
Phys. Rev. \textbf{108}, 1063 (1957).


\bibitem{zerilli}
F.~J.~Zerilli, ``Effective potential for even-parity Regge-Wheeler 
gravitational perturbation equations",
Phys. Rev. Lett. \textbf{24}, 737 (1970).


\bibitem{ckjpsl}
V.~Cardoso, R.~Konoplya, and J.~P.~S.~Lemos,
``Quasinormal frequencies of Schwarzschild black holes in 
anti-de Sitter space-times: A complete study on the asymptotic behavior",
Phys. Rev. D \textbf{68}, 044024 (2003);
arXiv:gr-qc/0305037 [gr-qc].


\bibitem{konoplya}
R.~A.~Konoplya,
``Massive vector field perturbations in the Schwarzschild background: 
Stability and quasinormal spectrum",
Phys. Rev. D \textbf{73}, 024009 (2006);
arXiv:gr-qc/0509026 [gr-qc].


\bibitem{redkov}
E.~M.~Ovsiyuk and V.~M.~Redkov,
``Spherical waves of spin-1 particle in anti de Sitter space-time",
Acta Phys. Pol. B \textbf{41}, 1247 (2010);
arXiv:1109.0387 [math-ph].

\bibitem{dolan}
J.~G.~Rosa and S.~R.~Dolan,
``Massive vector fields on the Schwarzschild spacetime:
Quasinormal modes and bound states",
Phys. Rev. D \textbf{85}, 044043 (2012);
arXiv:1110.4494 [hep-th].


\bibitem{Herdeiro2}
C.~Herdeiro, M.~O.~P.~Sampaio, and M.~Wang,
``Maxwell perturbations on asymptotically anti-de Sitter spacetimes: 
Generic boundary conditions and a new branch of quasinormal modes", 
Phys. Rev. D \textbf{92}, 124006 (2015);
arXiv:1510.04713 [gr-qc].




\bibitem{fhlc}
T.~V.~Fernandes, D.~Hilditch, J.~P.~S.~Lemos, and V.~Cardoso,
``Quasinormal modes of Proca fields in a Schwarzschild-AdS spacetime",
Phys. Rev. D \textbf{105}, 044017 (2022); arXiv:2112.03282 [gr-qc].



\bibitem{tf}
T.~V.~Fernandes, D.~Hilditch, J.~P.~S.~Lemos, and V.~Cardoso, ``Normal
modes of Proca fields in AdS spacetime",
Gen. Relativ. Gravit. \textbf{55}, 5 (2023); arXiv:2301.10248 [gr-qc].






\bibitem{burgess}
C.~P.~Burgess and C.~A.~Lutken,
``Propagators and effective potentials in anti-de Sitter space",
Phys. Lett. B \textbf{153}, 137 (1985);




\bibitem{ishibashi_kodama1}
H. Kodama and A. Ishibashi, ``A master equation for gravitational 
perturbations of maximally symmetric black holes in higher dimensions", 
Prog. Theor. Phys. {\bf 110}, 701 (2003);
arXiv:hep-th/0305147.

\bibitem{ishibashiwald}
A.~Ishibashi and R.~M.~Wald,
``Dynamics in nonglobally hyperbolic static space-times.
III. Anti-de Sitter space-time",
Classical Quantum Gravity \textbf{21}, 2981 (2004);
arXiv:hep-th/0402184 [hep-th].



\bibitem{natario}
J.~Nat\'ario and R.~Schiappa,
``On the classification of asymptotic quasinormal frequencies 
for $d$-dimensional black holes and quantum gravity",
Adv. Theor. Math. Phys. \textbf{8}, 1001 (2004);
arXiv:hep-th/0411267 [hep-th].


\bibitem{ortega}
A.~Lopez-Ortega,
``Electromagnetic quasinormal modes of $D$-dimensional black holes",
Gen. Relativ. Gravit. \textbf{38}, 1747 (2006);
arXiv:gr-qc/0605034 [gr-qc].



\bibitem{ishibashi_kodama2}
A. Ishibashi and H. Kodama, ``Perturbations and stability of 
static black holes in higher dimensions", 
Prog. Theor. Phys. Supp. {\bf 189}, 165 (2011);
arXiv:1103.6148 [hep-th].


\bibitem{Herdeiro}
C.~Herdeiro, M.~O.~P.~Sampaio, and M.~Wang,
``Hawking radiation for a Proca field in $D$-dimensions",
Phys. Rev. D \textbf{85}, 024005 (2012);
arXiv:1110.2485 [gr-qc].



\bibitem{ueda}
K.~Ueda and A.~Ishibashi,
``Massive vector field perturbations on extremal and near-extremal
static black holes",
Phys. Rev. D \textbf{97}, 124050 (2018);
arXiv:1805.02479 [gr-qc].




\bibitem{myers}
A.~Chodos and E.~Myers,
``Gravitational contribution to the Casimir energy in Kaluza-Klein theories",
Ann. Phys. (N.Y.) \textbf{156}, 412  (1984).


\bibitem{higuchi}
A.~Higuchi,
``Symmetric tensor spherical harmonics on the $N$ sphere and 
their application to the de Sitter group SO($N$,1)",
J. Math. Phys. \textbf{28}, 1553 (1987).



\bibitem{lovenumbers}
L.~Hui, A.~Joyce, R.~Penco, L.~Santoni, and A.~R.~Solomon,
``Static response and Love numbers of Schwarzschild black holes",
J. Cosmol. Astropart. Phys. JCAP 04 (2021) 052;
arXiv:2010.00593 [hep-th].




\bibitem{abramowitz}
M.~Abramowitz and I.~A.~Stegun, {\it
Handbook of Mathematical Functions
with Formulas, Graphs, and Mathematical Tables},
(Dover, New York, 1964). 


\end{thebibliography}
\end{document}